\documentclass[english, usenatbib]{mn2e}
\usepackage{epsfig}
\usepackage{fix2col}
\usepackage{lmodern}
\usepackage[T1]{fontenc}
\usepackage[latin9]{inputenc}
\usepackage[a4paper]{geometry}
\usepackage{units}
\usepackage{amsmath}
\usepackage{amssymb}


\usepackage{babel}
\usepackage{hyperref}
\usepackage{verbatim}

\title[Star-formation and Structural Properties]
      {The Relationship Between Star-formation Activity and Galaxy Structural Properties in CANDELS and a Semi-analytic Model}
      \author[Brennan et al.]{Ryan Brennan$^{1}$\thanks{E-mail:
          brennan@physics.rutgers.edu}, Viraj Pandya$^{2}$, Rachel
        S. Somerville$^{1}$, Guillermo Barro$^{3}$, \and Asa F. L. Bluck$^{4}$, Edward N. Taylor$^{5}$, Stijn Wuyts$^{6}$, Eric F. Bell$^{7}$, Avishai Dekel$^{8}$, \and Sandra Faber$^{3}$, Henry C. Ferguson$^{9}$, Anton M. Koekemoer$^{9}$, Peter Kurczynski$^{1}$, \and Daniel H. McIntosh$^{10}$, Jeffrey A. Newman$^{11}$, Joel Primack$^{12}$\\ $^{1}$Department of
        Physics and Astronomy, Rutgers, The State University of New
        Jersey, 136 Frelinghuysen Rd, Piscataway, NJ
        \\ $^{2}$Department of Astrophysical Sciences, Peyton Hall,
        Princeton University, Princeton, NJ \\ 
$^{3}$Department of Astronomy, University of California, Berkeley, CA \\
        $^{4}$Institute for Astronomy, Department of Physics, ETH Zurich, Wolfgang-Pauli-Strasse 27, Zurich, 8093, Switzerland \\
$^{5}$School of Physics, the University of Melbourne,
        Parkville, VIC 3010, Australia \\ $^{6}$Department of Physics, University of Bath, Claverton Down, Bath, BA2 7AY, UK \\ $^{7}$Department of Astronomy, University of Michigan, Ann Arbor, MI\\ 
$^{8}$Center for Astrophysical and Planetary Science, Racah Institute of Physics, The Hebrew University, Jerusalem 91904, Israel \\ $^{9}$Space Telescope Science Institute, 3700 San Martin Drive, Baltimore, MD \\ $^{10}$Department of Physics and Astronomy, University of Missouri-Kansas City, 5110 Rockhill Road, Kansas City, MO \\ $^{11}$Department of Physics and Astronomy, University of Pittsburgh and PITT-PACC, 3941 OHara St. Pittsburgh, PA \\$^{12}$Department of Physics, University of California at Santa Cruz, Santa Cruz, CA}

\begin{document}
\date{}

\maketitle

\label{firstpage}
\begin{abstract}

We study the correlation of galaxy structural properties with their
location relative to the SFR-$M_{*}$ correlation, also known as the
star formation ``main sequence'' (SFMS), in the CANDELS and GAMA
surveys and in a semi-analytic model (SAM) of galaxy formation. We
first study the distribution of median S{\'e}rsic index, effective
radius, star formation rate (SFR) density and stellar mass density in
the SFR-$M_{*}$ plane. We then define a redshift dependent main
sequence and examine the medians of these quantities as a function of
distance from this main sequence, both above (higher SFRs) and below
(lower SFRs). Finally, we examine the distributions of distance from
the main sequence in bins of these quantities. We find strong
correlations between all of these galaxy structural properties and the
distance from the SFMS, such that as we move from galaxies above the
SFMS to those below it, we see a nearly monotonic trend towards higher
median S{\'e}rsic index, smaller radius, lower SFR density, and higher
stellar density. In the semi-analytic model, bulge growth is driven by
mergers and disk instabilities, and is accompanied by the growth of a
supermassive black hole which can regulate or quench star formation
via Active Galactic Nucleus (AGN) feedback. We find that our model
qualitatively reproduces the trends described above, supporting a
picture in which black holes and bulges co-evolve, and AGN feedback
plays a critical role in moving galaxies off of the SFMS.
\end{abstract}


\section{Introduction}

Out to $z\sim3$, galaxies can be split into star-forming and quiescent
populations based on the bimodality observed in their colors and
derived star formation rates \citep{Baldry2004, Bell2004b,
  Brinchmann2004, Kauffmann2003, Strateva2001, Brammer2011,
  Ilbert2013}. When focusing specifically on the galaxies classified
as star-forming, a strong correlation is observed between the star
formation rate and stellar mass of galaxies at a fixed redshift (the
SFR-$\rm{M_{*}}$ correlation) \citep{Noeske2007, Daddi2007, Elbaz2007,
  Rodighiero2011}. This correlation is also sometimes referred to as
the ``star-forming main sequence'' (SFMS). This stands in contrast to
the less rigidly defined quiescent population, for which there is no such
strong correlation.

The SFR-$\rm{M_{*}}$ correlation can be defined by a
(redshift-dependent) normalization and slope, with a straight line in
log-log space providing a reasonable fit, although there is evidence
that the slope of the main sequence may flatten above a mass of
$\sim10^{10}M_{\odot}$ \citep{Whitaker2012, Whitaker2014}. It is still
unclear whether this flattening is simply due to the fact that more of
the stellar mass in high mass galaxies is likely to be in a
non-star-forming bulge component, as suggested by
\citet{Abramson2014} or \citet{Tacchella2015}, or whether there is something else going on. It
has also been suggested that the presence of non star-forming bulges
in star-forming galaxies may increase the scatter in the
SFR-$\rm{M_{*}}$ relation around the main sequence
\citep{Whitaker2015}. In any case, many studies have examined the
SFR-$\rm{M_{*}}$ correlation and found that it holds over at least
four orders of magnitude in mass and exists out to $z\sim6$ (see
\citet{Speagle2014} and references therein, as well as
\citet{Salmon2015}). The value of the slope in the SFR-$\rm{M_{*}}$
plane is measured to be $\sim1$ \citep{Rodighiero2011} and the
relationship has an intrinsic $1-\sigma$ scatter of only $\sim0.2-0.4$
dex \citep{Whitaker2012, Kurczynski2016}. In general, SFMS galaxies at
high redshift have much higher SFRs than galaxies on the main sequence
today \citep{Sobral2014}, and the evolution of the normalization of
the SFMS appears to be independent of galaxy environment
\citep{Peng2010}.

The small scatter of the SFR-$\rm{M_{*}}$ correlation leads us to
believe that galaxy evolution is dominated by relatively steady star
formation histories, rather than being highly stochastic and
bursty. This places constraints on the duty cycle of processes such as
galaxy mergers or disk instabilities, which may trigger starburst and
quenching events that drive galaxies above or below the main
sequence. Furthermore, observations show that since $z\sim 2$ there
has been a build-up of quiescent galaxies, while the mass
density of galaxies on the SFMS has remained relatively constant,
implying that galaxies are being moved \emph{off} of the SFMS into the
quiescent population, and remaining there permenantly or at least over
rather long timescales \citep{Bell2004b, Borch2006, Bell2007,
  Faber2007}. As the processes which move galaxies off of the main
sequence are often associated with morphological change, it is
interesting to examine the correlation between distance from the
SFR-$\rm{M_{*}}$ relation, or some other measure of quiescence, and
galaxy structural properties.

\citet[][B15]{Brennan2015} defined a redshift dependent SFMS by which to judge galaxies in order
to divide them into star-forming and quiescent populations. We
split the sSFR-S{\'e}rsic index plane into four quadrants in star-formation
activity and morphology: star-forming disk-dominated galaxies,
star-forming spheroid-dominated galaxies, quiescent disk-dominated
galaxies, and quiescent spheroid-dominated galaxies. After dividing
galaxies up, we examined the evolution of the fraction of galaxies in
each of these categories with redshift.  In order to constrain which
processes were responsible for moving galaxies between these different
categories, we did the same analysis on a sample of model galaxies
generated from the ``Santa Cruz'' semi-analytic model described in
\citet{Somerville2008} with updates as described in
\citet{Somerville2012} and \citet{Porter2014}. In addition to
prescriptions for the main physical processes believed to be important
for shaping galaxy properties (described below), the model includes
bulge formation due to mergers and disk instabilities, and concurrent
growth of supermassive black holes and AGN feedback, allowing us to
predict how model galaxies evolve in the SFR-S{\'e}rsic index plane.  The
SAM is a useful tool for studying the evolution of large populations
of galaxies, as it can generate large cosmologically representative
samples with modest computational resources, allowing us to
efficiently test the effects of various physical processes. In B15, we
found that our prescriptions for quenching and morphological
transformation were able to transform galaxies in a manner in
qualitative agreement with the observations as long as bulge growth
due to disk instabilities was included. Bulge growth due to mergers
and disk instabilities and subsequent AGN feedback produced roughly
the right fraction of galaxies in each of our four
subpopulations. Models in which bulge growth occured only due to
mergers did not produce as many spheroid-dominated galaxies as seen in
observations.

Our goal in this paper is to study the structural properties of model
galaxies \emph{continuously} across and off the main sequence, rather
than using the main sequence to sort our galaxies into bins based on
their SFRs and morphologies as in B15 and Pandya et al. (in
prep.). The latter explicitly examines galaxies with intermediate
star-formation and structural properties. We learned in B15 that our
model could broadly produce the right fractions of different types of
galaxies and the evolution of these fractions, and now we will examine
more closely if it can produce both ``typical'' main sequence
galaxies, as well as match how the structural properties of galaxies
change as they move farther from the main sequence. In this way, we
hope to continue to build our understanding of the physical processes
which drive the correlation between star formation, quenching, and
galaxy structural properties.

Many observational studies have examined the structure of galaxies
across the main sequence and come to several conclusions: 1) The main
sequence is made up of kinematically and morphologically disk-dominated galaxies which have the largest
radial sizes for their stellar masses \citep{Williams2010, Wuyts2011,
  Bluck2014, vanderwel2014b} (although it is true that the easy morphological distinction between disk-dominated and spheroid-dominated galaxies begins to break down at higher redshift, especially at high mass).  2) Galaxies lying above the main
sequence are often morphologically disturbed and seem to be undergoing
a starburst \citep{Wuyts2011, Elbaz2011, Salmi2012}. \citet{Elbaz2011}
suggests that some of these may also include heavily obscured AGN. Of course, morphological disturbance above the main sequence is not universal; see \citet{Barro2016}). 3)
Compact star-forming galaxies (cSFGs, as defined in \citet{Barro2013}) on or just below the main
sequence at z$\sim2$-3 suggest that bulge growth precedes quenching
\citep{Barro2014, Williams2014, Fang2015}. \citet{Fang2015} even found
that at z$\sim2$-3, cSFGs dominate the high mass end of the main
sequence.  4) Quiescence is almost always associated with a bulge
component or high central stellar mass density or velocity
dispersion\citep{Franx2008, Bell2012, Wake2012, Fang2013, Bluck2014, Lang2014, Woo2015, Teimoorinia2016}. Estimates of galaxy black hole masses derived from
central velocity dispersions also point to black hole mass being very
correlated with quiescence \citep{Bluck2016}. 

On the simulation side, \citet{Snyder2015} investigated the
relationship between optical morphology, stellar mass and star
formation rate for a sample of simulated galaxies from the Illustris
simulation \citep{Vogelsberger2014}. They found that their model,
which includes feedback from accreting supermassive black holes, was
able to produce the population of quiescent bulge-dominated galaxies
at $z\sim0$ needed to reproduce the distribution of observed
morphologies. \citet{Tacchella2016} examined how galaxies in the VELA
simulations \citep{Ceverino2014} oscillate around the main
sequence due to clumpy inflows and violent disk instabilities, leading
to compaction and minor quenching episodes.

We compare our model predictions with galaxies observed with the
Cosmic Assembly Near-infrared Deep Extragalactic Legacy Survey
(CANDELS; \citealt{Grogin2011, Koekemoer2011}) and the Galaxy and Mass
Assembly Survey (GAMA; \citealt{Driver2011}). In order to assure high
levels of completeness and robust measurements of structural
parameters, we consider only galaxies with stellar mass
$M_{*}$>$10^{10}M_{\odot}$, for both the models and observations. We
consider several structural properties, including S{\'e}rsic index,
size, stellar mass density and star-formation rate density. In Section
2, we describe the semi-analytic model in more detail and give a
summary of the observational data to which we will compare. In Section
3, we examine the distribution of structural properties across the
SFR-$\rm{M_{*}}$ plane, as in \citet[][hereafter W11]{Wuyts2011}. We
also examine how some quantities on which we currently do not have
direct observational constraints, such as bulge-to-total luminosity
ratio, black hole mass, and dark matter halo mass, vary across this
plane. We next consider how these structural properties change as a
function of linear distance from the main sequence, again as studied
by W11. In Section 4, we investigate the distribution of distances
from the main sequence in bins of galaxy structural properties
following the analysis of \citet{Bluck2014}. A secondary goal of the
paper will be to compare our results to those of W11 and
\citet{Bluck2014}, the inspirations for several of our plots, where
appropriate, and we discuss this in Section 5, along with a comparison
between our model predictions and some other theoretical predictions
in the literature. In Section 5 we also discuss what our model tells
us about the universe in the cases where our model and the
observations agree, and what the universe is telling us about our
model in the cases where they don't. We summarize our results and
conclude in Section 6.

\section{Semi-Analytic Model and Observational Data}

\subsection{The Semi-Analytic Model}

In this work, we use the same semi-analytic model as was used in B15,
which was first presented in \citet{Somerville1999} and
\citet{Somerville2001} and updated in \citet[][S08]{Somerville2008},
\citet[][S12]{Somerville2012} and \citet[][P14]{Porter2014}. This model has been shown to produce populations of galaxies that are in good agreement with observations; for comparisons with several statistical galaxy properties, see S08 and S12, as well as \citet{Lu2014} and \citet{Somerville2015} for the evolution of the stellar mass function out to high redshift. A detailed look at the size-mass relation will appear in Somerville et al. (in prep.). As noted
in B15, this model includes prescriptions for the hierarchical growth
of structure, heating and cooling of gas, star formation, stellar
population evolution, supernova feedback, chemical evolution of the
interstellar medium (ISM) and intracluster medium (ICM) due to
supernovae, and AGN feedback, as well as starbursts and morphological
transformation due to mergers between galaxies and disk instabilities
in isolated galaxies. We briefly summarize these processes below. For
a more detailed description of the processes governing quenching and
morphological transformation, see B15. For a more detailed
description of the model in general, see S08 and P14. We assume a
$\Lambda$CDM cosmology ($\Omega_{m}$=0.307, $\Omega_{\Lambda}$=0.693,
h=0.678) and a \citet{Chabrier2003} initial mass function. We have
adopted a baryon fraction of 0.1578. Our cosmology is consistent with
the Planck 2013 results
\citep{Planck2014} and was chosen to match that of the
Bolshoi Planck simulation \citep{RodriguezPuebla2016}.

We use CANDELS mock lightcones (Somerville et al. in prep.) extracted
from the Bolshoi Planck dark-matter N-body simulation
\citep{Klypin2011, Trujillo2011, RodriguezPuebla2016}. The ROCKSTAR
algorithm of \citet{Behroozi2013} is used to identify dark matter
halos. Merger trees for each halo in the light cone are constructed
using the method of \citet{SK:1999}, updated as described in S08. For
our lowest redshift bin, we use a snapshot from the Bolshoi volume as
opposed to the lightcone, which at that redshift represents a very
small volume.

When dark matter haloes merge, the central galaxy of the largest
progenitor becomes the new central galaxy, while all other galaxies
become satellites.  Satellite galaxies are able to spiral in and merge
with the central galaxy, losing angular momentum to dynamical friction
as they orbit.  The merger time-scale is estimated using a variant of
the Chandrasekhar formula from \citet{Boylan2008}. Tidal stripping and
destruction of satellites as described in S08 are also included.

Before the universe is reionized, each halo has a hot gas mass equal
to the virial mass of the halo times the universal baryon fraction.
The collapse of gas into low mass haloes is suppressed after
reionization due to the photoionizing background. We assume the
universe is fully reionized by $z=11$ and use the results of
\citet{Gnedin2000} and \citet{Kravtsov2004} to model the fraction of
baryons that can collapse into haloes of a given mass following
reionization.

When dark matter haloes collapse or are involved in a merger that at
least doubles the mass of the progenitors (a 1:1 merger), the hot gas
is shock-heated to the virial temperature of the new halo. The rate at
which this gas can cool is determined by a simple spherical cooling
flow model (see \citet{Somerville2008} for details) which approximates
the transition from ``cold flows'', where cold gas streams into the
halo along dense filaments without being heated, to ``hot flows'',
where gas is shock heated on its way in, forming a diffuse hot gas
halo before cooling \citep{Birnboim2003, Keres2005, Dekel2006}. In
this way, virial shock heating is included in our SAM, although many
studies show that this effect alone is not enough to produce the
observed population of massive quiescent galaxies \citep[][and
  references therein]{SomervilleDave2015}.

\subsubsection{Star-formation, Bulge-formation and AGN Feedback}

There are two modes of star formation in the model: a ``normal'' mode
that occurs in isolated disks and a ``starburst'' mode that occurs as
a result of a merger or internal disk instability, discussed
below. The normal mode follows the Schmidt-Kennicutt relation
\citep{Kennicutt1998} and assumes that gas must be above some fixed
critical surface density (the adopted value here is $6\,
M_{\odot}/\rm{pc}^{2}$) in order to form stars.

Newly cooled gas collapses to form a rotationally supported disk, the
scale radius of which is estimated based on the initial angular
momentum of the gas and the profile of the halo. We assume that
angular momentum is conserved and that the self-gravity of the
collapsing baryons causes the inner part of the halo to contract
\citep{Blumenthal1986,Flores1993,Mo1998}.

Exploding supernovae and massive stars are capable of depositing
energy into the ISM, which can drive outflows of cold gas from the
galaxy.  We assume that the mass outflow rate is proportional to the
SFR and decreases with increasing galaxy circular velocity, in
accordance with the theory of ``energy-driven'' winds
\citep{Kauffmann1993b}. Some ejected gas is removed from the halo
completely, while some is deposited into the hot gas reservoir of the
halo and is eligible to cool again. The gas that is driven from the
halo entirely is combined with the gas that has been prevented from
cooling by the photoionizing background and may later reaccrete back
into the halo. The fraction of gas which is retained by the halo
versus the amount that is ejected is a function of halo circular
velocity as decribed in S08.

Heavy elements are produced by each generation of stars, and chemical
enrichment is modelled simply using the instantaneous recycling
approximation. For each parcel of new stars $dm_{*}$, a mass of metals
$dM_{\rm{Z}}=ydm_{*}$ is also created, which is immediately mixed with
the cold gas in the disk. The yield $y$ is assumed to be constant
and is treated as a free parameter. Supernova driven winds act to
remove some of this enriched gas, depositing a portion of the created
metals into the hot gas or outside of the halo.

Spheroids can be created by mergers or disk instabilities. Mergers
between galaxies are assumed to remove angular momentum from the stars
and gas in the disk and drive material towards the center, building up
a spheroidal component. In our model, this component is formed
instantaneously. The size of the spheroid is determined by the stellar
masses, sizes and gas fractions of the progenitors with the help of
hydrodynamical binary merger simulations (see P14). This is a
relatively recent addition to the model which gives much more accurate
galaxy sizes. Velocity dispersions are also computed as described in
P14. Mergers also trigger a starburst, the efficiency of which depends
on the gas fraction of the central galaxy and the mass ratio of the
two progenitors. The parameterization for the efficiency and timescale
of the burst is based on hydrodynamical simulations of mergers between
disk galaxies \citep{Hopkins2009a}. Stars formed as part of the burst
are added to the spheroidal component, as are 80\% of the stars from
the merging satellite galaxy. The other 20\% are assumed to be
distributed in a diffuse stellar halo. We note here that the merger
fraction in our model has been shown to be consistent with
observational estimates in \citet{Lotz2011}.

Disk material can also be converted into a spheroidal component as a
result of internal gravitational instabilities.  A pure disk without a
dark matter halo is very unstable to the formation of a bar or bulge,
while massive dark matter haloes tend to stabilize a thin, cold
galactic disk \citep{Ostriker1973,Fall1980}. When the ratio of dark
matter mass to disk mass falls below a critical value, the disk can no
longer support itself and material collapses into the inner regions of
the galaxy \citep{Efstathiou1982}. Here we adopt an avenue for bulge
growth due to disk instability, based on a Toomre-like stability
criterion, which can be found in P14 or B15. As with mergers, the
creation of the bulge component is instantaneous. This is another
relatively recent addition to the model, but an important one as was
shown in B15. In that work, we toggled the disk instability on and
off. As we found that the model including disk instabilities was more
successful in reproducing the morphological mix of galaxies seen in
CANDELS and the local Universe, we focus almost exclusively on that
one here.

Galaxies are initially seeded with a massive black hole of
$10^{4} \, M_{\odot}$ \citep{Hirschmann2012}. This black hole is allowed
to grow by accretion and merging with other black holes, particularly
as a result of galaxy mergers and disk instabilities. The prescription
for this growth is described in B15 and in more detail in
\citep{Hirschmann2012}. In the case of mergers, the accretion rate is
based on hydrodynamical binary merger simulations \citep{Hopkins2006,
  Hopkins2007}. For disk instabilities, the black hole is allowed to
feed on some fraction of the mass which is moved to the spheroidal
component, here $10^{-3}$, as in \citet{Hirschmann2012}. In these
cases, the black hole enacts feedback in the form of radiatively
efficient, or ``quasar'' mode, AGN activity. It is also able to feed
and effect feedback in the ``radio'' or ``maintenance'' mode, during
which it feeds via Bondi-Hoyle accretion from the galaxy's hot halo
\citep{Bondi1952}.

\subsection{Computing S{\'e}rsic Index and composite size for model galaxies}

In this work we compare the structural properties of model
  galaxies to those of observed galaxies. Our main basis of comparison
  is the S{\'e}rsic index; although, as mentioned above,
  disk-dominated and spheroid-dominated galaxies at high redshift
  become less morphologically distinct, the S{\'e}rsic index should
  still provide us with information about whether we are dealing with
  an extended or more compact galaxy. Our model directly computes the
bulge luminosity, total luminosity, bulge radius and disk radius for
each galaxy, allowing us to compute the bulge-to-total H-band flux
ratio and bulge radius-to-disk radius ratio. The bulge radius and disk
radius are recorded by the model as the 3D half-mass radius of stars
in the bulge and the 3D scale radius of stars and cold gas in the disk
respectively. For our high redshift galaxies (z>0.5), we convert these quantities to projected rest-frame
V-band half-light radii (the projection done according to
  \citet{Prugniel1997}) in order to use the stellar mass and redshift
  dependent wavelength correction provided by \citet{vanderwel2014} to
  get observed frame H-band sizes to go with our H-band bulge-to-total
  ratio (and to match the observed H-band S{\'e}rsic indices and sizes from CANDELS). The sizes of our low redshift model galaxies are left in the rest-frame V-band, which should be comparable to the r-band from which the structural properties of GAMA galaxies are derived. We then utilize a lookup table generated by creating
  synthetic galaxies that are composites of $n=1$ (disk) and $n=4$
  (spheroid) components, and then fitting a single component
  S{\'e}rsic profile to the synthetic image (see \citet{Lang2014} for
  details). The table is parameterized in terms of bulge-to-total
  ratio and the ratio of the effective radii of the bulge and disk
  components. The output is an effective S{\'e}rsic index and
  effective radius for the composite system. The table contains
  discrete values so we use a 2D interpolation.  The S{\'e}rsic index
  and effective radius that we derive here are light-weighted, in
  contrast with the stellar mass weighted quantities used in B15, and
  should provide a more accurate comparison to the S{\'e}rsic indices
  and sizes derived from light for our observed sample. However, we
  note that we do not attempt to include the effects of dust
  attenuation in our light-weighted quantities. We find that adopting
  these light-weighted quantities does not qualitatively change our
  results relative to B15, but does result in a significant
  improvement in the agreement between our models and the
  observations.

\subsection{Observational Data}
\subsubsection{High Redshift: CANDELS}

Our high redshift dataset (spanning $0.5<z<2.5$) consists of
observations taken as part of the Cosmic Assembly Near-infrared Deep
Extragalactic Legacy Survey (CANDELS; \citealt{Grogin2011,
  Koekemoer2011}). The CANDELS data span five different fields and in
this work we use data from all five: COSMOS (Nayyeri et al. (in
prep.)), GOODS-N (Barro et al. (in prep.)), GOODS-S \citep{Guo2013},
EGS (Stefanon et al., in prep.), and UDS \citep{Galametz2013} (see
these references for details about data processing and catalog
creation for each of the CANDELS fields.)
With this
multiwavelength data we are able to study the star formation
properties and structure of galaxies out to $z\sim2.5$ at high
resolution.

We make use of data catalogs generated by several previous
  studies. Here we give a very brief overview of the derivation of
physical parameters which applies generally to all of the CANDELS
fields (for more details see B15 and Pandya et al. (in prep.)). For a
given field, the template-fitting method TFIT \citep{Lee2012,
  Laidler2007} was used to merge datasets of different wavelengths
with different resolutions in order to construct the observed-frame
multi-wavelength photometric catalog. The Bayesian framework of
\citet{Dahlen2013} was used to derive photometric
redshifts. Spectroscopic redshifts are used where available and
reliable. 3D-HST grism redshifts are used for GOODS-S galaxies where
available \citep{Morris2015}. The EAZY code (\citealt{Brammer2008} and
Kocevski et al. (in prep.)) was used to fit templates to the
observed-frame SEDs in order to derive rest-frame photometry. Several
independent codes, such as FAST \citep{Kriek2009}, were used to derive
stellar masses under fixed assumptions, but allowing for some
variation of assumed star formation histories. We assume the
following: \citet{BC2003} stellar population synthesis models,
\citet{Chabrier2003} initial mass function, exponentially declining
star formation histories, solar metallicity and a \citet{Calzetti2001}
dust attenuation law. A ladder of SFR indicators prescribed in
\citet{Barro2011} and W11 is used to derive star formation rates for
galaxies in each field as described in B15 (see Section 2.2.1 of that
work). Finally, structural parameters were derived using GALFIT
\citep{Peng2002}, fitting to the HST/WFC3 F160W H-band images using a
one-component S{\'e}rsic model as described in \citet{vanderwel2012}.

We make the following selection cuts on our data: stellar mass >
$10^{10}M_{\odot}$ (to ensure completeness) and GALFIT quality flag=0
(to ensure good fits and robustness of our galaxy morphologies). We
cut at a stellar mass of $10^{10}M_{\odot}$ for continuity with our
low-redshift GAMA sample, which starts to become incomplete below this
mass range. Because of this, we employ a relatively conservative mass
cut throughout this work.

\subsubsection{Low Redshift Sample: GAMA}

At low redshift CANDELS probes a very small volume, so we supplement
with observations from Data Release 2 (DR2) of the Galaxy and Mass
Assembly survey (GAMA;\citealt{Liske2015}). Our low redshift range
spans $0.005<z<0.12$, sometimes referred to as $z=0.06$ in the
text. GAMA has an area of 144 square degrees and goes two magnitudes
deeper (r<19.8 mag) than SDSS while maintaining high spectroscopic
completeness ($\gtrsim98\%$). GAMA also has a rich supplementary
multi-wavelength dataset \citep{Liske2015}. The backbone of GAMA is
deep optical spectroscopy with the Anglo-Australian Telescope (AAT),
while its multi-wavelength catalogs are bolstered by collaborations
with several other independent surveys (for a review, see
\citet{Driver2011}).

Again we make use of derived properties generated by previous
  work. Bulk flow-corrected redshifts are adopted from
\citet{Baldry2012} and rest-frame photometry and stellar masses were
derived from SED fitting as described in \citet{Taylor2011}. GAMA's
high spectroscopic completeness allows the derivation of
$H\alpha$-based star formation rates from extinction-corrected
$H\alpha$ line luminosities. Structural properties of GAMA galaxies
are provided via multi-band measurements using GALFIT
\citep{Peng2002}. We adopt the structural fits in the r-band so as to
analyze the structural properties of GAMA galaxies in the same band in
which they were selected (as with the H-band for CANDELS galaxies).

We also employ the following selection cuts as we did with our CANDELS
data: stellar mass > $10^{10}\,M_{\odot}$ (again to ensure our
sample is complete), and GALFIT quality flag = 0.

\subsection{Defining the Main Sequence}

We define the main sequence much as we did in B15, although this time
we use log(SFR) instead of log(sSFR). As described in B15, we decide
to define our own stellar mass and redshift dependent main sequence
line that is determined by the mean star formation rates of
galaxies. The star formation rates of observed galaxies are
systematically slightly higher than those of model galaxies, so this
line is calculated separately for observed and model galaxies. While
this already means that our model galaxies are not behaving exactly as
observed galaxies, we do not think it impedes our goal of examining
galaxy properties relative to the main sequence; as we judge
quantities in this paper as a function of distance from the main
sequence line, we don't expect the disparity in absolute star
formation rates to affect our results.

We calculate the average star
formation rate of galaxies with stellar masses between $10^{9}$ and
$10^{9.5}M_{\odot}$ in time bins in order to measure the baseline main
sequence star formation rate across cosmic time, as we do not believe
galaxies of this mass will be contaminated by objectively quiescent
galaxies. In the models we use only central galaxies. We then
calculate the main sequence slope by measuring the change in the mean
log(SFR) between stellar masses of $10^{9}$ and $10^{9.5}M_{\odot}$. In
each redshift bin, we use the mean low mass SFR and derived slope to
define a mass-dependent main sequence line. While the SFR-$\rm{M_{*}}$
correlation is known to have some dispersion, in order to judge
distance from the main sequence, we define it with a single line, as
has been done in W11 and \citet{Bluck2014}. We also note that while
the observed main sequence slope is known to flatten toward higher
stellar mass \citep{Whitaker2012, Schreiber2015, Lee2015}, here we extrapolate the slope
derived for lower stellar mass galaxies to higher mass. We do this
following the interpretation that the decrease in slope at higher
stellar mass is due to the higher probability the galaxies of larger
mass are already starting to quench and move off of the main
sequence. Here, we try to define a more ``pristine'' version of the
main sequence, which we expect on theoretical grounds based on the
fact that models without quenching have an unbroken linear
SFR-$\rm{M_{*}}$ correlation (see \citet{Renzini2015} for another
alternative to defining an unbiased main sequence). Throughout this
work, distance from the main sequence is given in units of log(SFR).

\subsection{Evolution of Star-forming Galaxies vs. Quiescent Galaxies in the SAM}

\begin{figure*}
\epsfig{file=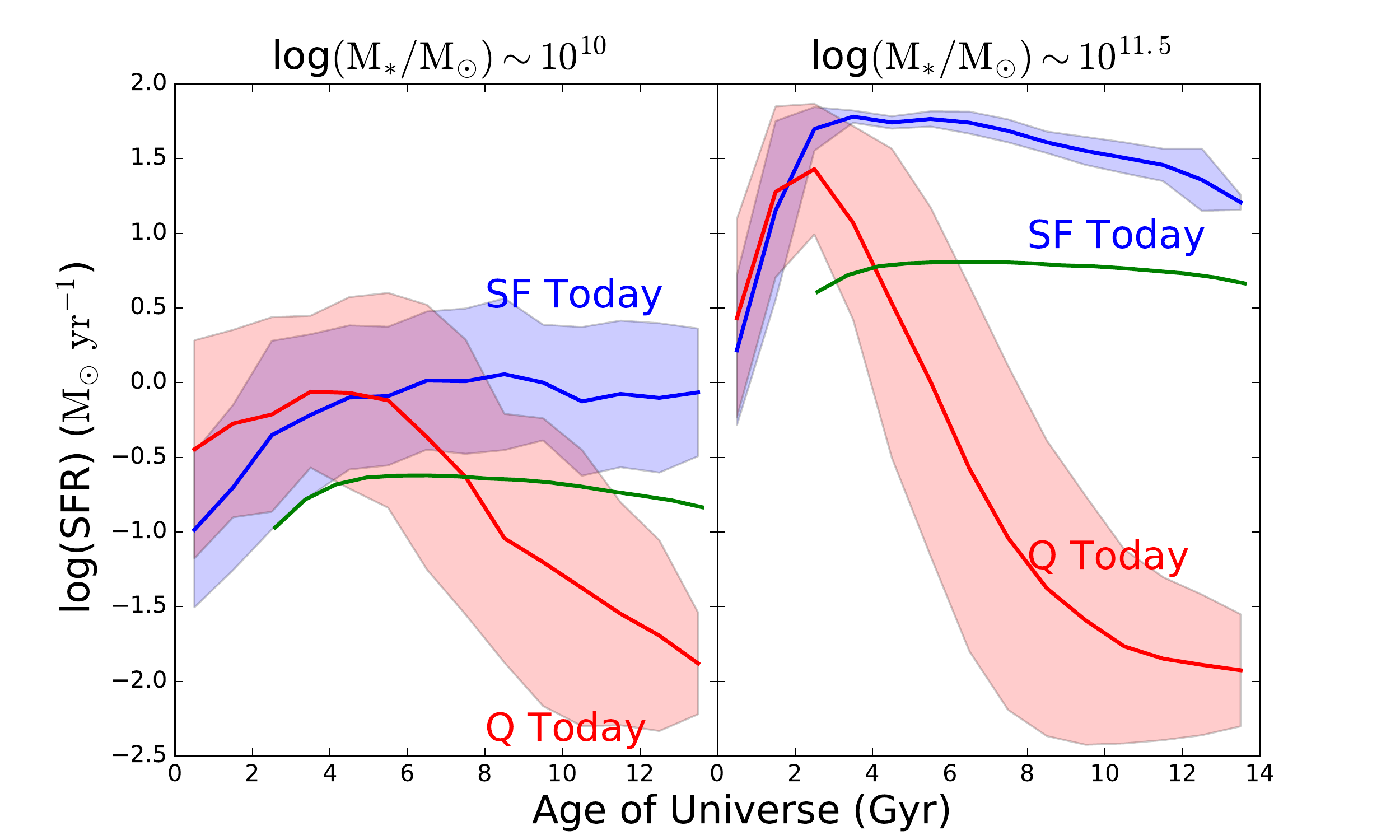, width=1.0\textwidth}
\caption{Evolution of the mean SFR for model galaxies that are
  star-forming or quiescent in the present day, split into two mass
  bins. The blue lines indicate star-forming galaxies at $z=0$, while
  red indicates quiescent galaxies at $z=0$. The shaded regions
  correspond to the $1-\sigma$ scatter in SFR for each of the
  curves. The green lines indicate the time-dependent star formation
  cut off line, below which galaxies are considered quiescent at a
  given age of the universe. Left panel: Galaxies with stellar masses
  $\sim10^{10}M_{\odot}$ at $z=0$. Right panel: Galaxies with stellar
  masses $\sim10^{11.5}M_{\odot}$ at $z=0$. We see the main sequence
  of star formation manifested in the higher SFRs of more massive
  galaxies. We also see evidence of downsizing in the SFRs of more
  massive galaxies, which peak earlier than those of less massive
  galaxies. This is true for both galaxies that are star-forming
  today, and for quiescent galaxies which peak at early times before
  falling below our quiescence threshold.}
  {\label{sfrevol}}
\end{figure*}

Figure \ref{sfrevol} shows the evolution of the average star formation
rate of galaxies from our model with cosmic time. The blue lines
correspond to galaxies that are considered star-forming at $z=0$
according to the prescription described in B15 (meaning their SFRs are
greater than 25\% of main
sequence SFR described above), while the red lines correspond to
quiescent galaxies at $z=0$ (with SFR < 25\% of the main sequence
SFR). Galaxies have been split into two mass bins at z=0, with final stellar masses
$\sim10^{10}M_{\odot}$ ($10^{9.9}$-$10^{10.1}M_{\odot}$) or $10^{11.5}M_{\odot}$ ($10^{11.4}$-$10^{11.6}M_{\odot}$), representing the two
ends of the mass range we are considering. We see evidence for the
SFR-$\rm{M_{*}}$ correlation in the higher SFR for star-forming
galaxies of higher stellar mass (right panel) versus that of the lower
stellar mass galaxies in the left panel. We also see an overall
decrease in the SFRs of massive galaxies with cosmic time after an
early peak at $\sim2$-3 Gyrs. The SFRs of the less massive galaxies
are only now beginning to decrease. We see the same type of behavior
for the quiescent galaxies, with the higher mass quiescent galaxies
exhibiting an earlier and stronger peak.

The scatter in SFR for quiescent galaxies is in general larger than
that for star-forming galaxies because the mechanism that leads to the
most intense quenching, AGN feedback, is associated with significant
mass growth due to the major and minor mergers that trigger it. This
is not as apparent in the low mass panel, but only because we have
artificially put a floor at log(SFR)=-2.0. Otherwise, the mean quiescent
SFR became much less well-behaved.

This difference in average star formation histories between
star-forming and quiescent galaxies is indicative of the SFMS at work
in our model. As discussed in B15, and mentioned above, the
SFR-$\rm{M_{*}}$ correlation in our model does not behave exactly as
that observed in the universe; while the slope and normalization of
our model main sequence is not quite the same as the observed
SFR-$\rm{M_{*}}$ correlation, we do reproduce a relationship between
SFR and stellar mass. Galaxies tend to stay near this sequence until
something happens to move them off of it, and the diversity of
processes responsible, as well as the varying severity of these
processes, leads to the larger spread in average star formation
histories of galaxies that are quiescent today. Later, we will examine
different galaxy properties as a function of distance from this
star-forming main sequence, but first we will look at how different
galaxy properties are distributed in the SFR-$\rm{M_{*}}$ plane.

\section{Distribution of Properties in the Star formation Rate-Stellar Mass Plane}

Here we examine how the median S{\'e}rsic index, effective radius, SFR
density, and stellar mass density vary across the SFR-$M_{*}$ plane
for both our model and the observations.  We note again here that for
the rest of this work we impose a floor on log(SFR) so that all
log(SFR)<-2.0 are set equal to -2.0. This is mainly to deal with
quiescent model galaxy SFRs which would be far below the plots
otherwise.

\subsection{Number Density in SFR-$\rm{M_{*}}$ Plane}

\begin{figure*}
  \epsfig{file=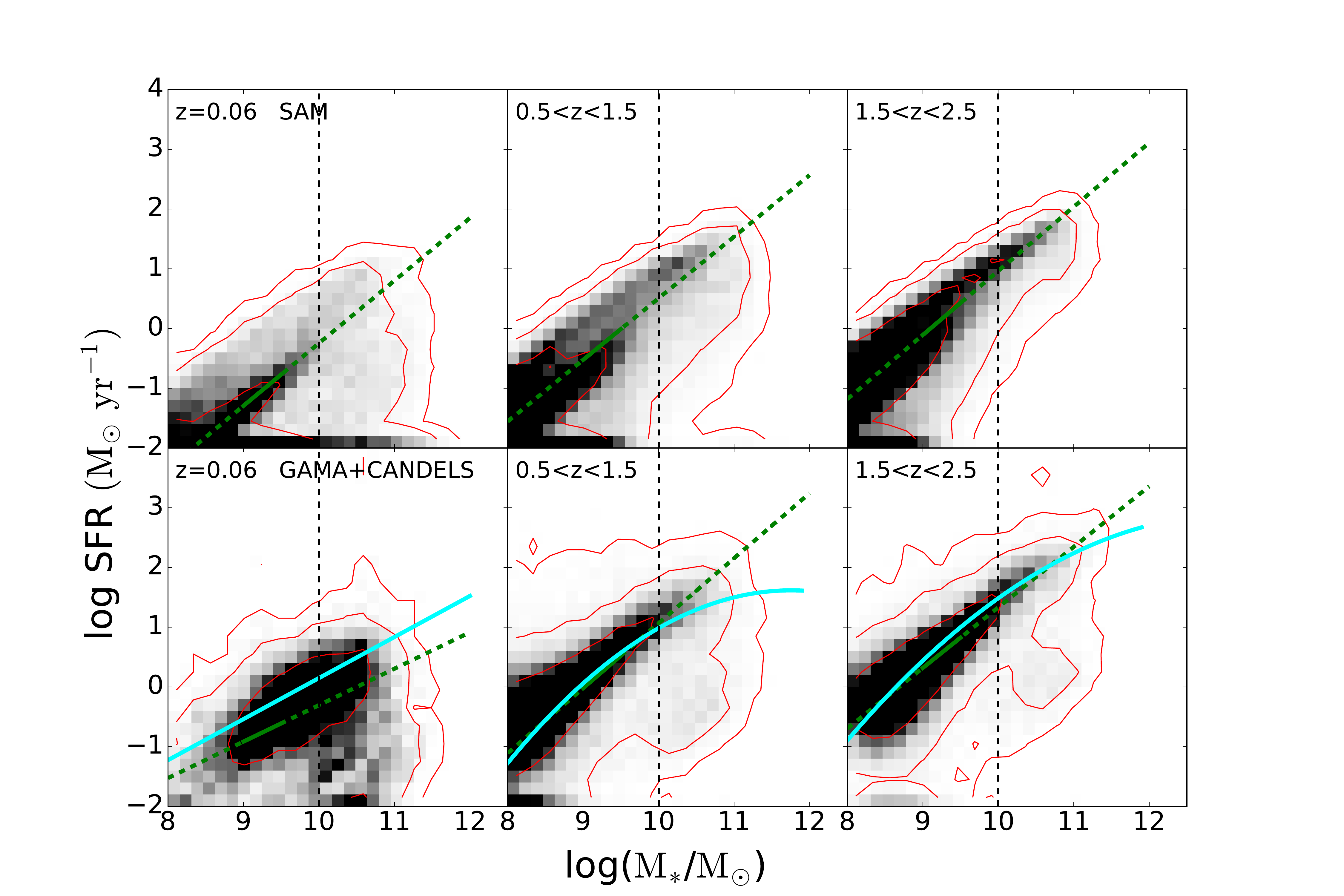, width=0.9\textwidth}
  \caption{The distribution of model (top) and observed (bottom)
  galaxies with stellar mass >$10^{8}M_{\odot}$ in the plane of SFR vs
  stellar mass in the redshift bins $z\sim0.06$ (left panels),
  $0.5<z<1.5$ (middle panels), and $1.5<z<2.5$ (right panels). The
  greyscale indicates population density with contours overlaid in
  red. The green lines show our fits to the main sequence of star
  formation, which are based on the mass range $10^{9}$ to
  $10^{9.5}M_{\odot}$ (solid green) and extrapolated to higher and
  lower mass (dashed green). The cyan lines indicate the main sequence
  fits found in \citet{Whitaker2012} (lowest redshift bin) and
  \citet{Whitaker2014} (two higher redshift bins). The fits in
  \citet{Whitaker2014} are for smaller redshift bins, so we averaged
  the coefficients of the fits that fell within our larger bins. We
  see, as mentioned above, where the main sequence slope becomes more
  shallow at higher stellar mass. The dashed black lines show the
  stellar mass cut we use for the rest of this work. We note that the
  normalization and slope of the SFMS is slightly different in the
  models and in the observations, which is why we fit them
  separately. In addition, we note that the distributions of SFR for
  quiescent galaxies are quite different in the models and
  observations. We discuss this further in the main text.}

  {\label{distgrad}}
\end{figure*}

In Figure \ref{distgrad}, we show the distribution of galaxies in the
SFR vs stellar mass plane. The number density is shown in greyscale
with contours overlaid in red. We also show the main sequence fits we
derive for both the model and observations in the three redshift bins
of interest, as well as comparisons with the main sequence derived in
\citet{Whitaker2012} and \citet{Whitaker2014}. We see immediately
where the GAMA survey begins to become incomplete below a stellar mass
of $10^{10}M_{\odot}$, which is why we have cut at this mass. We find
that the distribution of galaxies is somewhat different in the model
than in the observations. At all redshifts, most quiescent galaxies in
our models have SFR that are below our floor value $\log(SFR)=-2.0$,
while in the observations there is a cloud of galaxies with SFR that
are low enough to qualify them as `quiescent' but well above our floor
value.

This may be due to limitations in our modeling of gas inflows and AGN
feedback (for example, we may not resolve short timescale rejuvenation
events), or it could be due to the difficulty of obtaining accurate
observational estimates of SFR for quiescent galaxies (while there is
no explicit floor on detected SFRs in CANDELS or GAMA, the errors at
low absolute SFR can become quite large and a natural floor is set
based on the upper limits of detection in the photometric band used to
derive the SFR).  Despite this difference, we see that the main
sequence fits seem reasonable given the underlying distributions and
continue with our analysis, although we will remark throughout when it
seems this underlying difference is responsible for deviations between
our model and the observations. We will also discuss possible reasons
for this difference in our Discussion section.

\subsection{S{\'e}rsic Index in SFR-$\rm{M_{*}}$ Plane}

\begin{figure*}
\epsfig{file=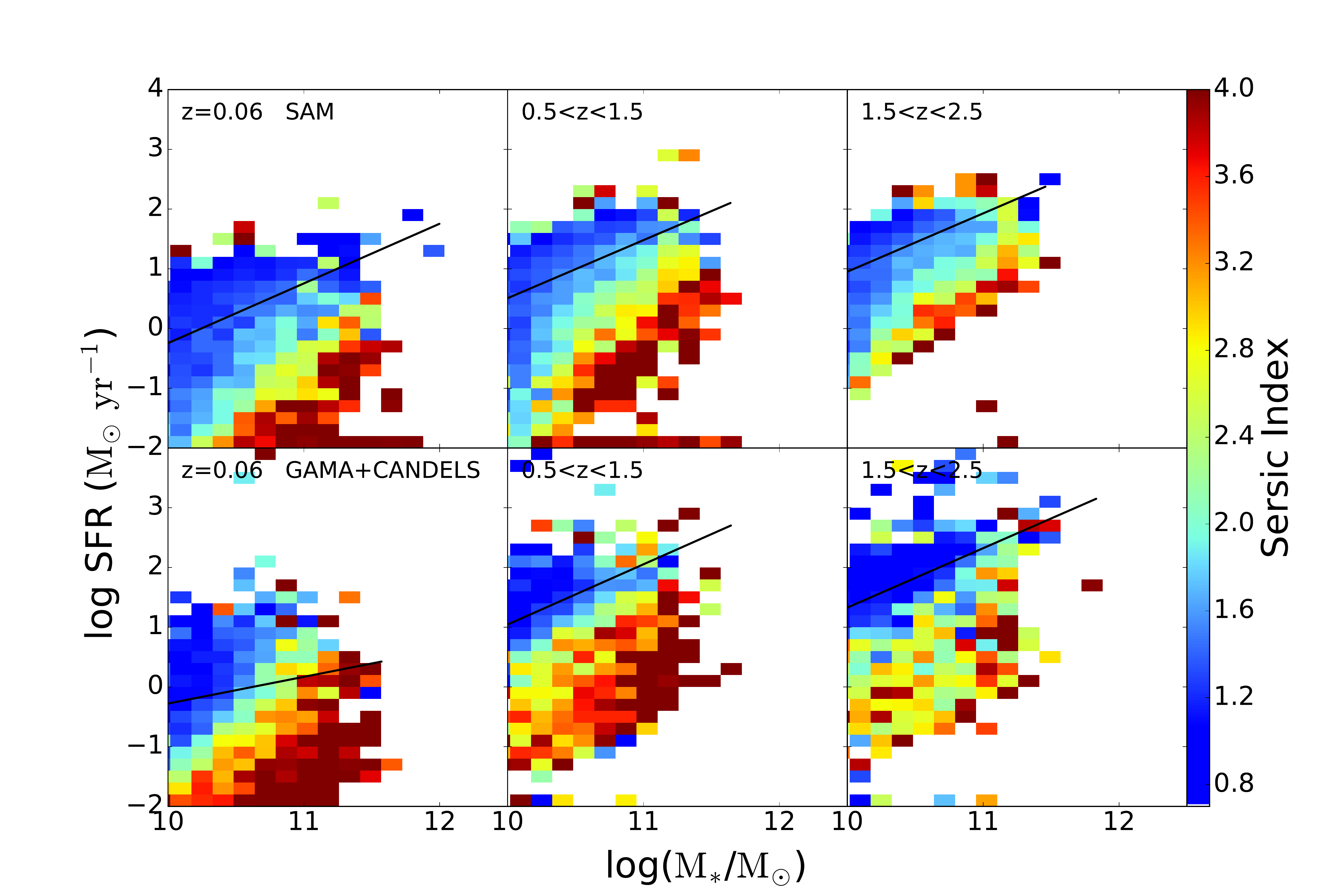, width=0.9\textwidth}
\caption{Distribution of median light-weighted S{\'e}rsic index in the
  SFR-$\rm{M_{*}}$ plane for (top) model galaxies and (bottom)
  observed galaxies in three redshift bins. The black lines indicate
  the star-forming main sequence fits. We find good qualitative
  agreement between the model predictions and the observations,
  although our model does not exactly reproduce the distribution of
  structural properties across the main sequence. In addition, massive
  high S{\'e}rsic-index ($n\sim4$) galaxies are more strongly quenched
  in our models than in the observations.
}
{\label{ngrad}}
\end{figure*}

In Figure \ref{ngrad}, we explore the distribution of S{\'e}rsic index
in the SFR-$\rm{M_{*}}$ plane by examining a color map of the median
S{\'e}rsic index in bins of SFR and $M_{*}$ (as in the analysis of
W11, to which we compare directly in Section \ref{sec:wuyts}). The top
panel shows this distribution for galaxies from our model and the
bottom panel shows galaxies from the GAMA and CANDELS surveys. We have
estimated the 1-$\sigma$ uncertainty on the median S{\'e}rsic index in
each observational bin due to uncertainties in the estimates of galaxy
properties in our observational sample. As in B15, we use quoted
uncertainties in S{\'e}rsic index and effective radius, an assumed
uncertainty of 0.25 log(SFR) for star-formation rates, and the
redshift-dependent stellar mass uncertainty of
\citet{Behroozi2013b}. The uncertainty, dn, in almost all bins and
across all redshifts is only $\sim0.0-0.3$, except for at high
redshift for low SFR galaxies, where dn$\sim1.0$. In our lowest
redshift bin, there is also a small patch of low SFR massive
(>$10^{11}M_{\odot}$) bins with dn$\sim2.0$. With this in mind we see
that the model and observational distributions are qualitatively quite
similar, although there are a few key differences.

Both the model and observations exhibit a pocket of high S{\'e}rsic
index at low SFR and high mass, although this trend is more pronounced
in the observations, especially in the two lower redshift bins. As
noted before, more galaxies in the models ``pile up'' at SFRs below our
floor value than in the observations, and these galaxies primarily
have high S{\'e}rsic index ($n\sim4$) characteristic of very
spheroid-dominated galaxies. In the observations, high-S{\'e}rsic index
(spheroid-dominated) galaxies are predominantly quiescent, but have
higher SFRs than their model counterparts. In addition, the quiescent
population is dominated by galaxies with higher
S{\'e}rsic index in the observations than in the models (see also
Figure \ref{ndist}).

Both the observations and model exhibit a smattering of high
S{\'e}rsic index galaxies along the top edge of the SFR-$\rm{M_{*}}$
distribution, above the main sequence, although this is more apparent
in the observations. In our models, we know that galaxies like these are star bursting as the result of a merger and appear as
bulge-dominated. However, we see that many of the highly star forming
galaxies above the main sequence in our model appear instead to be
disk-dominated. The difference is especially apparent in the middle
redshift bin, where some of the most massive, star-forming galaxies
appear to have very strong disks. This is also true \emph{on} the main
sequence, where there appears to be a sharper transition to
bulge-dominated systems along the observed main sequence (at
$\sim10^{11}M_{\odot}$ in our lowest redshift bin) than we see in the
model.

\begin{figure*}
\epsfig{file=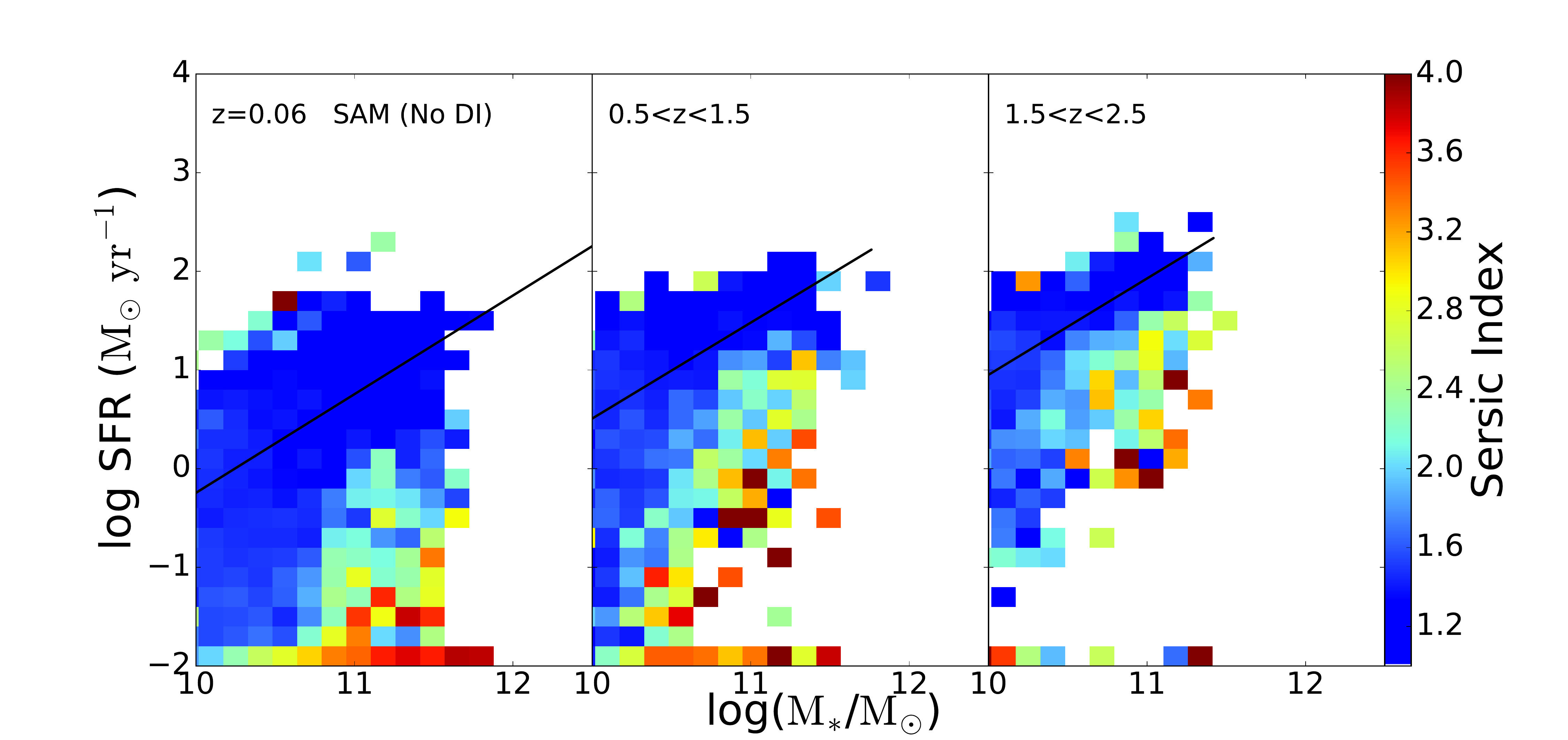, width=0.9\textwidth}
\caption{Distribution of S{\'e}rsic index in the SFR-$\rm{M_{*}}$
  plane for model galaxies in three redshift bins, in a version of the
  model that does not include bulge growth due to disk
  instabilities. The black lines indicate the star-forming main
  sequence fits. Here we see how important the disk instability
  mechanism is to building bulges in our model galaxies. Without it,
  we have very few bins with a median S{\'e}rsic index $\gtrsim3.5$.}
        {\label{nsgrad}}
\end{figure*}

In Figure \ref{nsgrad}, we show the distribution of S{\'e}rsic index
across the SFR-$\rm{M_{*}}$ plane for our model without the
prescription for bulge growth via disk instability. We see here how
important the disk instability is in producing bulge-dominated
galaxies. Without it we have very few truly bulge-dominated systems,
even at low redshift far below the main sequence. Our main sequence is
also completely dominated by disk galaxies, even at the high
mass end, unlike in the observed sample. The distribution of galaxies in the SFR-$\rm{M_{*}}$ plane as compared with Figure \ref{ngrad} is relatively unchanged; the disk instability is much more important for building bulge componenets than it is for quenching galaxies (see also B15, where it is shown that the quiescent fraction of galaxies changes very little between the two models, while the spheroid-dominated fraction changes by a significant amount.)

\subsection{Sizes and Surface Densities in SFR-$\rm{M_{*}}$ Plane}

\begin{figure*}
  \epsfig{file=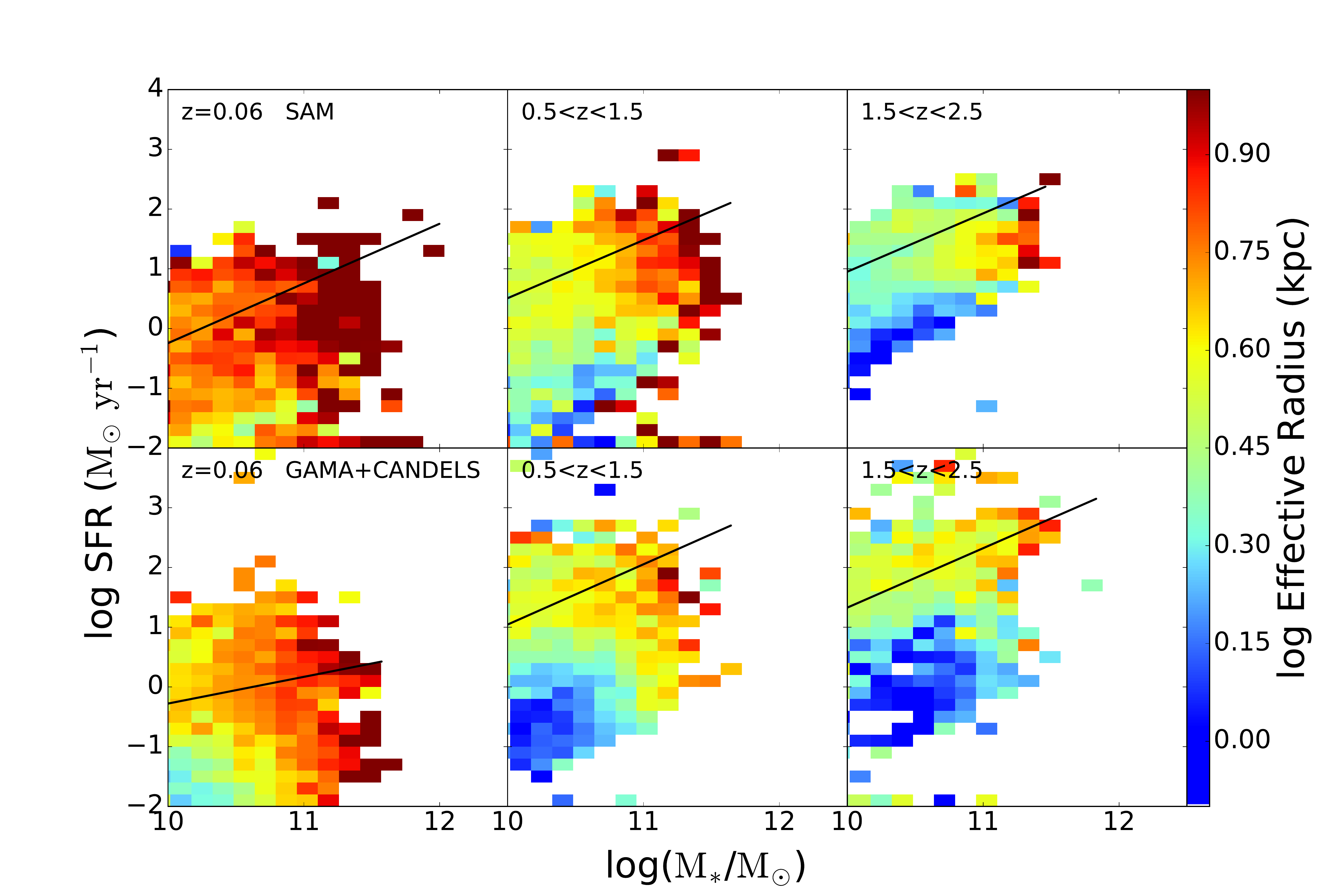, width=0.9\textwidth}
  \caption{Distribution of log median effective radius in the SFR-$\rm{M_{*}}$ plane for (top) model galaxies and (bottom) observed galaxies in three redshift bins. The black lines indicate the star-forming main sequence fits. The agreement between model and observations is qualitatively quite good, although at low redshift, our model produces galaxies that are too large}.
  {\label{regrad}}
\end{figure*}

Figure \ref{regrad} is the same as Figure \ref{ngrad}, but for log
effective radius. We calculate the uncertainty in each bin again as we
did for S{\'e}rsic index and find that in general it is only
$\sim0.05$ dex. In our lowest redshift bin at very high stellar mass,
the uncertainty can grow to be $\sim0.5$ dex, but this affects very
few bins. Again, the models qualitatively match the observations, although our model galaxies at low redshift tend to be too large. The
main features are that there is a clear sequence from compact to
extended galaxies from left to right, simply reflecting the size-mass
relation. There is no clear correlation between size and location in
the SFR-$\rm{M_{*}}$ plane for galaxies that are near the SFMS (see
also Shanahan et al., in prep.). However, galaxies that are below the
SFMS (quiescent galaxies) are more compact at almost every mass than
SF galaxies (although at our highest masses, even galaxies below the
main sequence tend to be quite extended; we will return to this in Section
\ref{sec:msdist}, as well as the issue of our large low redshift galaxies). These observational trends are well known
\citep[see e.g.][and references therein]{vanderwel2014b} and our
models qualitatively reproduce them. We discuss the quantitative
comparison between our model predictions and observational results in
more detail in Section \ref{sec:msdist}.

\begin{figure*}
  \epsfig{file=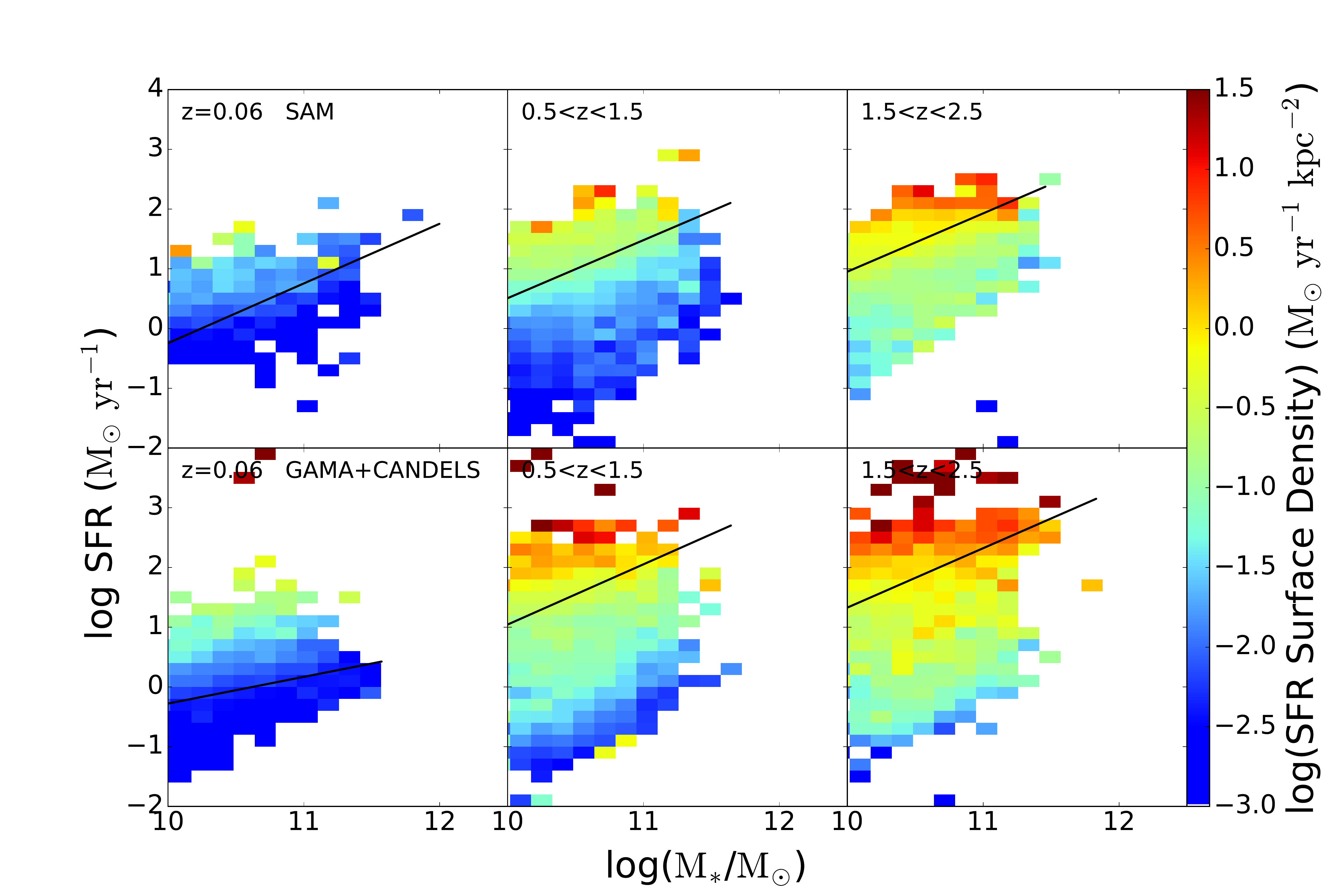, width=0.9\textwidth}
  \caption{Distribution of median SFR density in the SFR-$\rm{M_{*}}$
    plane for (top) model galaxies and (bottom) observed galaxies in
    three redshift bins. The main difference between the model and the
    observations in all redshift bins is the absence of the highest
    SFR density systems in the model as compared with the
    observations. This is due to the on average slightly larger radii
    of the model galaxies above the main sequence, where the most
    concentrated observed galaxies are.}  {\label{siggrad}}
\end{figure*}

Figure \ref{siggrad} is the same as Figures \ref{ngrad} and
\ref{regrad}, but now looking at the distribution of median SFR
surface density, $\Sigma_{\rm{SFR}}$, defined as $SFR$/$2\pi r^{2}$,
where $r$ is the effective radius. The uncertainty on these median
values is generally less than $\sim0.5$ dex. Here again we see very
good qualitative agreement between our models and the observations,
the biggest difference being the high density bins sitting above the
main sequence that are much more pronounced in the observations than
in the model. Whereas at
$z=0.06$, the model has no bins with a median
log($\Sigma_{\rm{SFR}}$)$\gtrsim1.0$ even at high redshift, the
observations show several high SFR bins with log($\Sigma_{\rm{SFR}}$)
as large as 1.5 all the way down to low redshift. This may reflect
limitations in our modeling of starburst systems.

\begin{figure*}
  \epsfig{file=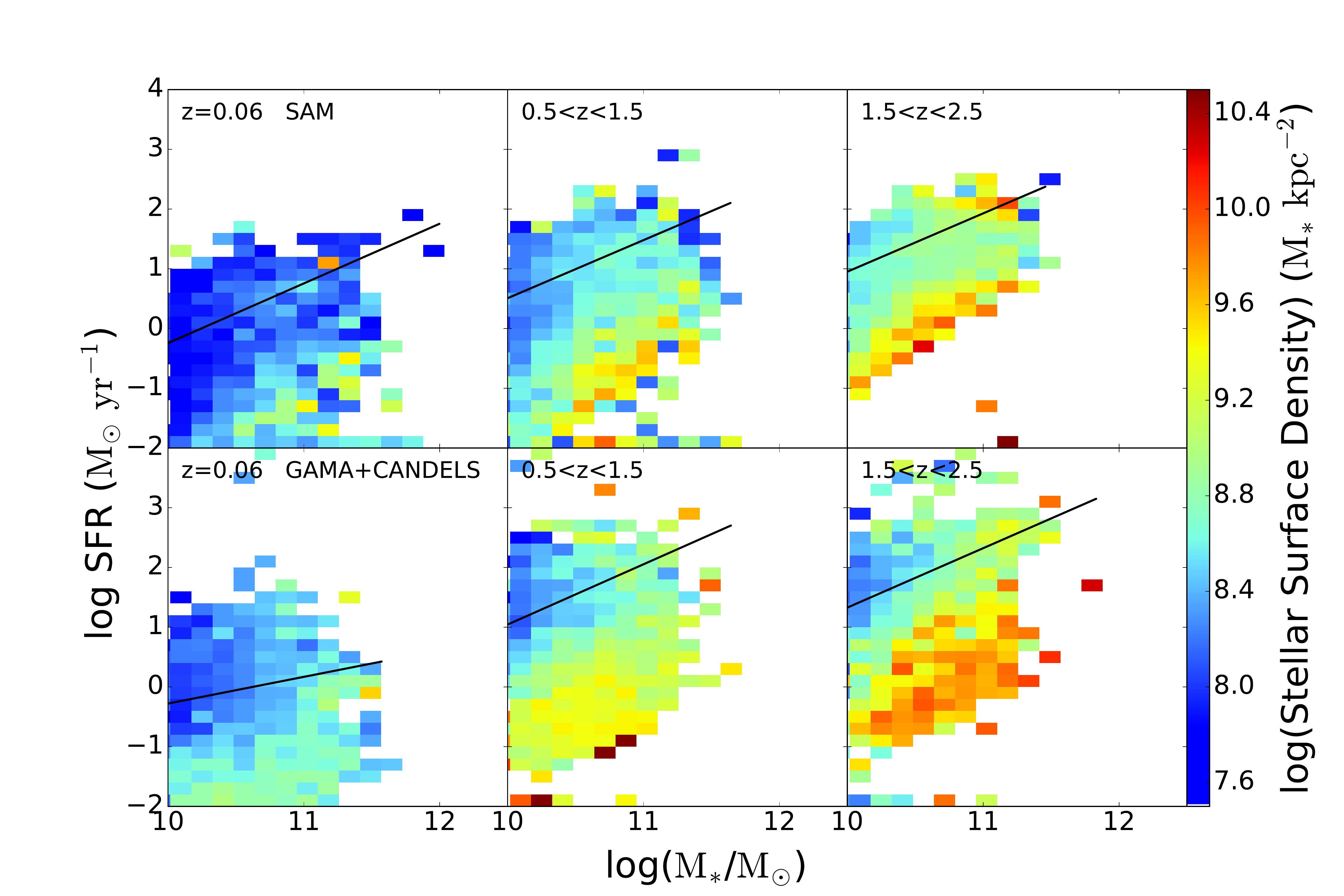, width=0.9\textwidth}
  \caption{Distribution of median stellar mass density in the
  SFR-$\rm{M_{*}}$ plane for (top) model galaxies and (bottom)
  observed galaxies in three redshift bins. The black lines indicate
  the star-forming main sequence fits. The qualitative agreement
  between the models and observations is very good, although the
  models do not reproduce as prominent a population of high surface
  density, quiescent galaxies in the highest redshift bin as seen in
  the observed distribution.}

  {\label{sigstargrad}}
\end{figure*}

Figure \ref{sigstargrad} shows the distribution of the median stellar
mass surface density, $\Sigma_{\rm{M_{*}}}$, defined as $M_{*}$/$2\pi
r^{2}$ in the SFR-$M_{*}$ plane. The uncertainties on the median
values here are less than $\sim1.0$ dex. We find that our agreement is
very good in the lowest redshift bin. At higher redshifts, we see more
compact systems in the observations, mainly below the main sequence at
high stellar mass, than we produce in our model. This is most
noticable in our highest redshift bin. The most compact systems are
those with $n\sim4$ in the quiescent cloud in Figure \ref{ngrad}.

\subsection{Model-only Properties in the SFR-$\rm{M_{*}}$ Plane}

\begin{figure*}
\epsfig{file=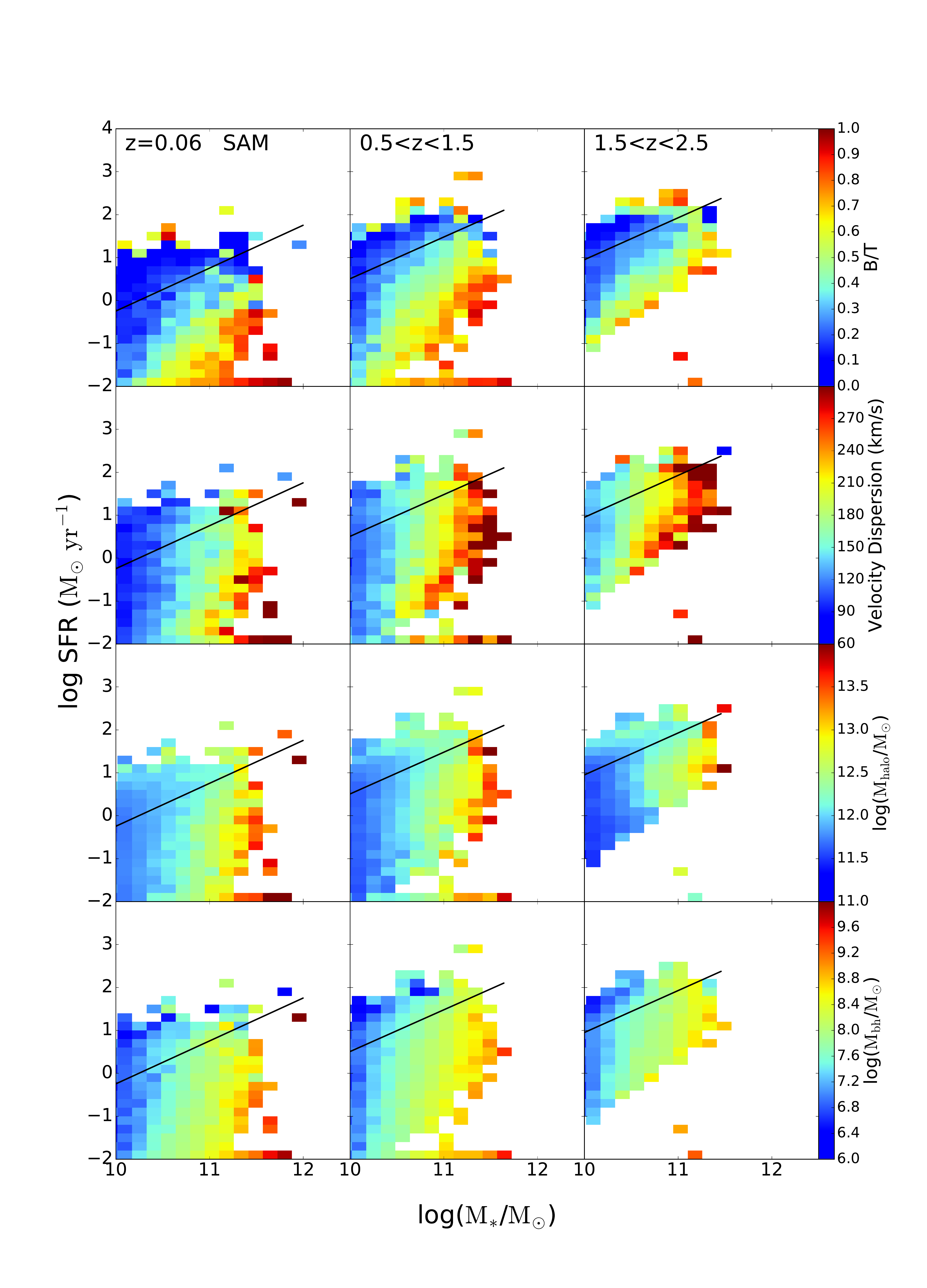, width=0.9\textwidth}
\caption{Distribution of (from top to bottom) bulge-to-total
  luminosity ratio in observed F160W, central velocity dispersion,
  halo mass, and black hole mass in the SFR-$\rm{M_{*}}$ plane for
  model galaxies in three redshift bins. The black lines indicate the
  star-forming main sequence line fits. Here we see the behavior of
  galaxy parameters which are native to our model, rather than the
  derived quantities we need to compare with observations. B/T and
  S{\'e}rsic index track each other very well. The other three
  quantities are strongly correlated with stellar mass.}
        {\label{modelgrad}}
\end{figure*}

In Figure \ref{modelgrad}, we look at the distribution of some
properties which are predicted for our models, but for which we do not
currently have direct observational constraints. However, all of these
quantities can in principle be observationally constrained. From top
to bottom, these are bulge-to-total luminosity ratio (in observed
F160W), bulge velocity dispersion, dark matter halo mass and black
hole mass. The diagrams look extremely similar for all of these. As
seen in previous studies, stellar mass is strongly correlated with all
of these quantities, which also have significant correlations with one
another. From this analysis, it is not possible to conclusively
determine which property is the most fundamental causal factor in
driving galaxy quiescence. We investigate this in more detail in a
future work.

\section{Distance from the Main Sequence}
\label{sec:msdist}
In order to be a bit more quantitative, we now examine the medians of
the quantities investigated in the previous section as a continuous
function of distance from the main sequence. We define $\Delta$SFR as
log(SFR)-log(SFR$_{MS}$), where log(SFR$_{MS}$) is the main sequence
SFR for a galaxy's stellar mass and redshift. A $\Delta$SFR of $\sim0$
indicates galaxies on the main sequence, while a positive or negative
$\Delta$SFR indicates galaxies above (with a higher SFR than) or below (with a lower SFR than) the main sequence,
respectively. The shaded region represents the distribution of the
25th-75th percentiles. We also include 1-$\sigma$ error bars derived
the same way as the uncertainties on the median quantities in the last
section. We set a floor for $\Delta$SFR at a value of -3 dex, below
which there are very few galaxies in either the models or the
observations. We also employ somewhat larger bins towards lower
$\Delta$SFR to combat very low number statistics.

\begin{figure*}
\epsfig{file=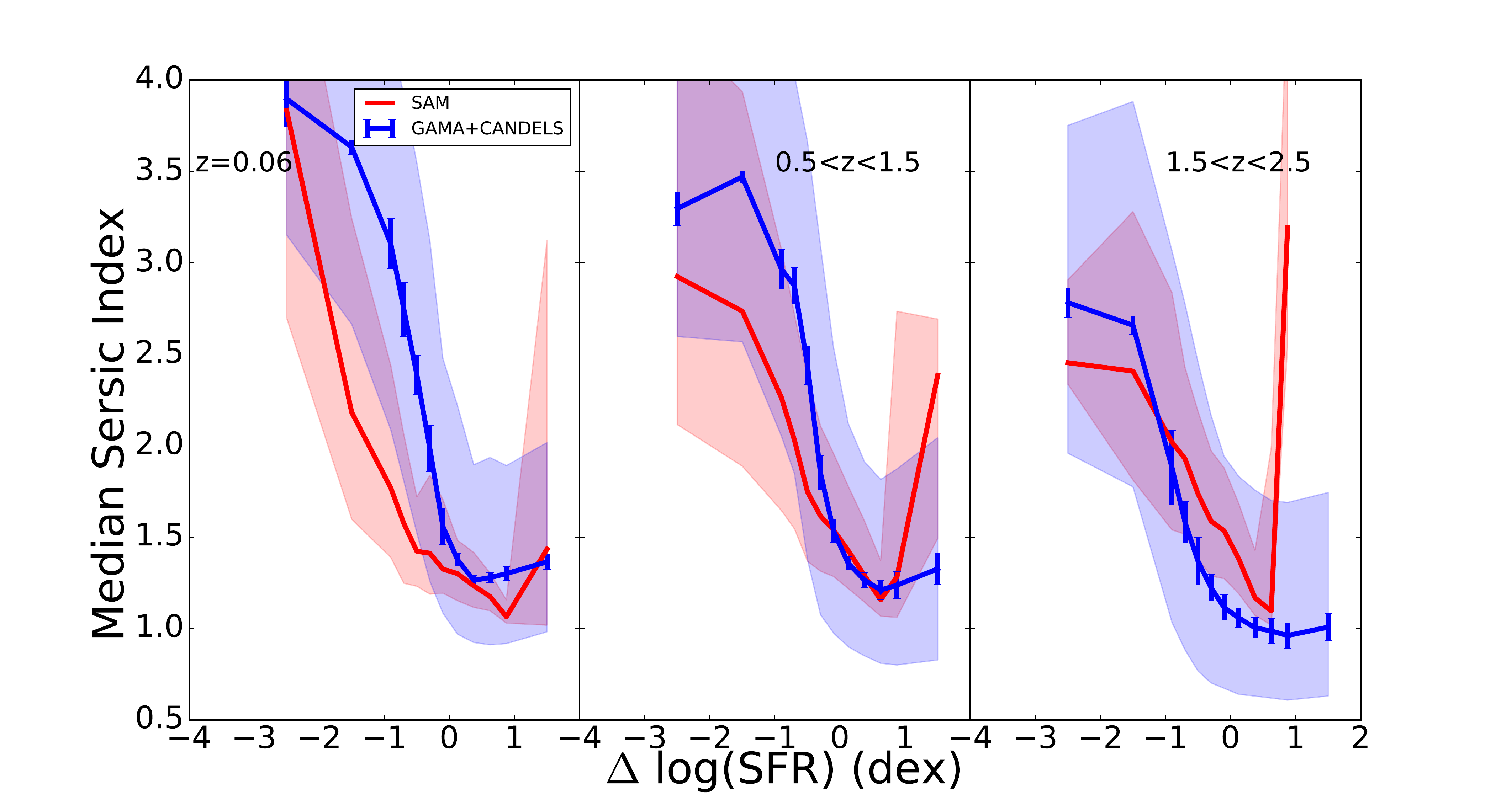, width=0.9\textwidth}
\caption{Median S{\'e}rsic index as a function of vertical distance
  from the fitted star-forming sequence for model galaxies (red) and
  observed galaxies (blue). The shaded region covers the 25th-75th
  percentiles of S{\'e}rsic index and the observations also have
  $1-\sigma$ error bars reflecting the uncertainties in galaxy
  parameter estimation. Below the SFMS, both the model and the
  observations exhibit an increase in S{\'e}rsic index with
  increasing distance from the SFMS.
%
}  {\label{ndist}}
\end{figure*}

Figure \ref{ndist} shows median S{\'e}rsic index as a function of
distance from the main sequence. In both the models and the
observations, we see that the SFMS is dominated by galaxies with low
S{\'e}rsic index (1.0-1.5), demonstrated by the minima of both the red
and blue curves in all redshift bins near $\Delta SFR=0$ (recall that
the intrinsic width of the SFMS is $\sim0.2-0.4$ dex). In the highest
redshift bin, the SFMS population has slightly \emph{lower} median
S{\'e}rsic index (closer to a pure $n=1$ exponential) in the
observations, while in the models the median S{\'e}rsic index in this
regime remains similar to the lower redshift bins. The trend towards
increasing S{\'e}rsic index with decreasing $\Delta SFR$ seen in the
observations is qualitatively reproduced in the models, as already
noted, but the region below the SFMS ($\Delta SFR <0$) is dominated by
galaxies with higher values of S{\'e}rsic index in the observations,
at least in the two lower redshift bins. In the highest redshift bin,
the models produce fewer quiescent galaxies than are seen, as already
noted and discussed (also in B15). For the observations, there is a
very slight upturn in median S{\'e}rsic index in the starburst regime
of the SFMS ($\Delta$ SFR $\gtrsim 0.6$; e.g. \citet{Rodighiero2011})
in the two lowest redshift bins. In the models, the highest
$\Delta$SFR bin is dominated by the few very highly star-forming,
newly bulge-dominated systems, resulting in the large spike seen in
all three redshift bins. These objects are rare in the model and
subject to large statistical fluctuations in our relatively small
samples, leading to the spikes as opposed to the gradual upturn of the
observations. We also might expect the upturn in the observations to be larger (as seen in W11), but if the starburst is triggered by processes that
cause morphological disturbance (such as mergers or disk
instabilities), they are likely to have been excluded from our
observational sample by our GALFIT quality cut.

\begin{figure*}
\epsfig{file=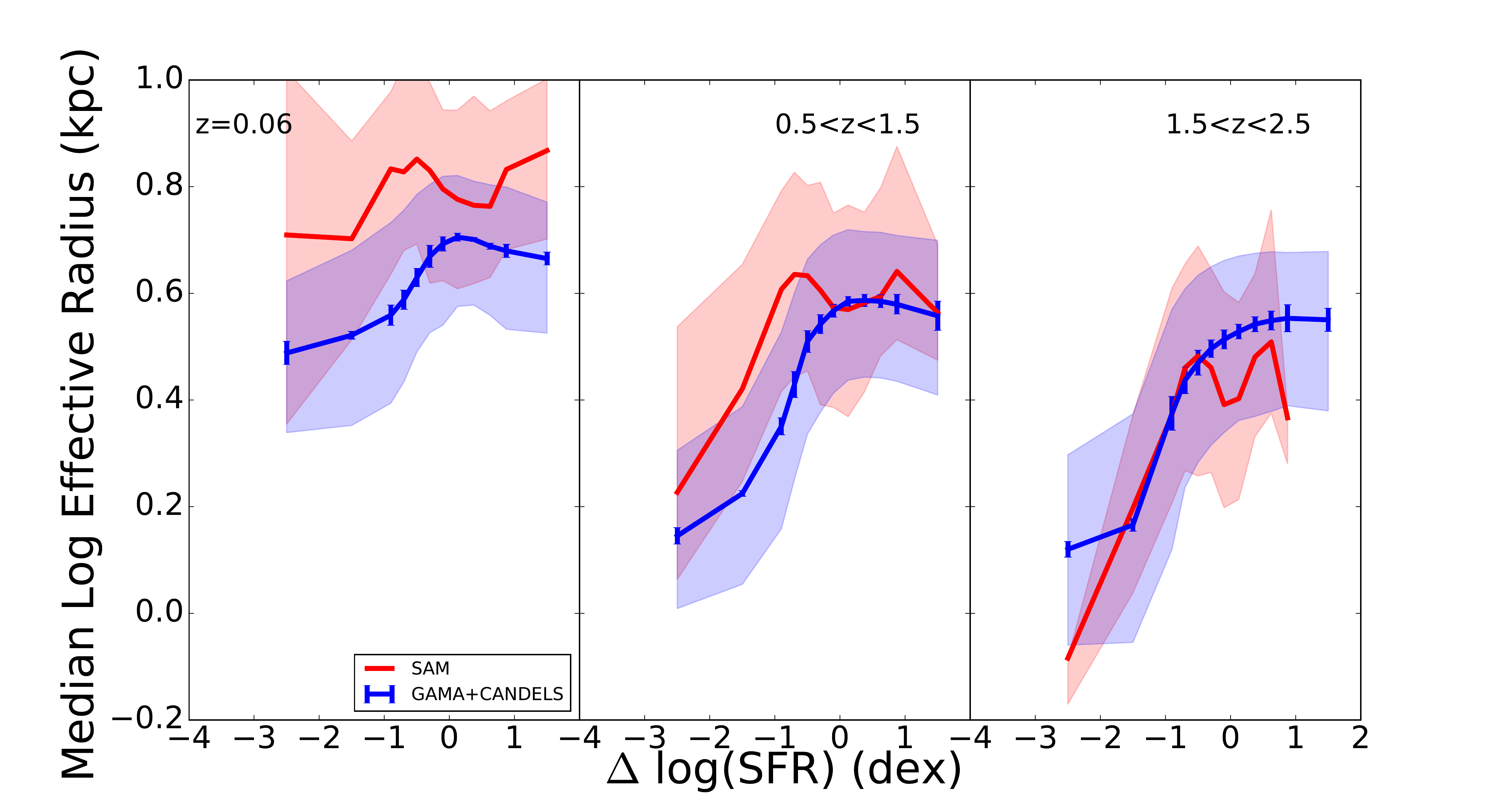, width=0.9\textwidth}
\caption{Median effective radius as a function of vertical distance
  from the fitted star-forming sequence for model galaxies (red) and
  observed galaxies (blue). The shaded region covers the 25th-75th
  percentiles of effective radius and the observations also have
  $1-\sigma$ error bars reflecting the uncertainty in galaxy parameter
  estimation. At low redshift, our model galaxies tend to be too large.}  {\label{redist}}
\end{figure*}

In Figure \ref{redist}, we see that our model in general produces
galaxies whose sizes are in rough agreement with their observational
counterparts in our two higher redshift bins, although in our
lowest redshift bin, the model tends to produce galaxies
which are too large regardless of distance from the main sequence. We
also see that in the model, galaxies just above and below the main
sequence tend to be slightly larger than galaxies directly on the main
sequence; see the Discussion for more details.  In our lowest redshift
bin, and to a lesser extent in our middle redshift bin, the galaxies furthest below the main sequence tend to be especially large compared to observed galaxies; this will also be discussed later. In the observations,
the largest galaxies live on the main sequence, with radial size
decreasing monotonically below the main sequence with increasing
distance from it.


\begin{figure*}
\epsfig{file=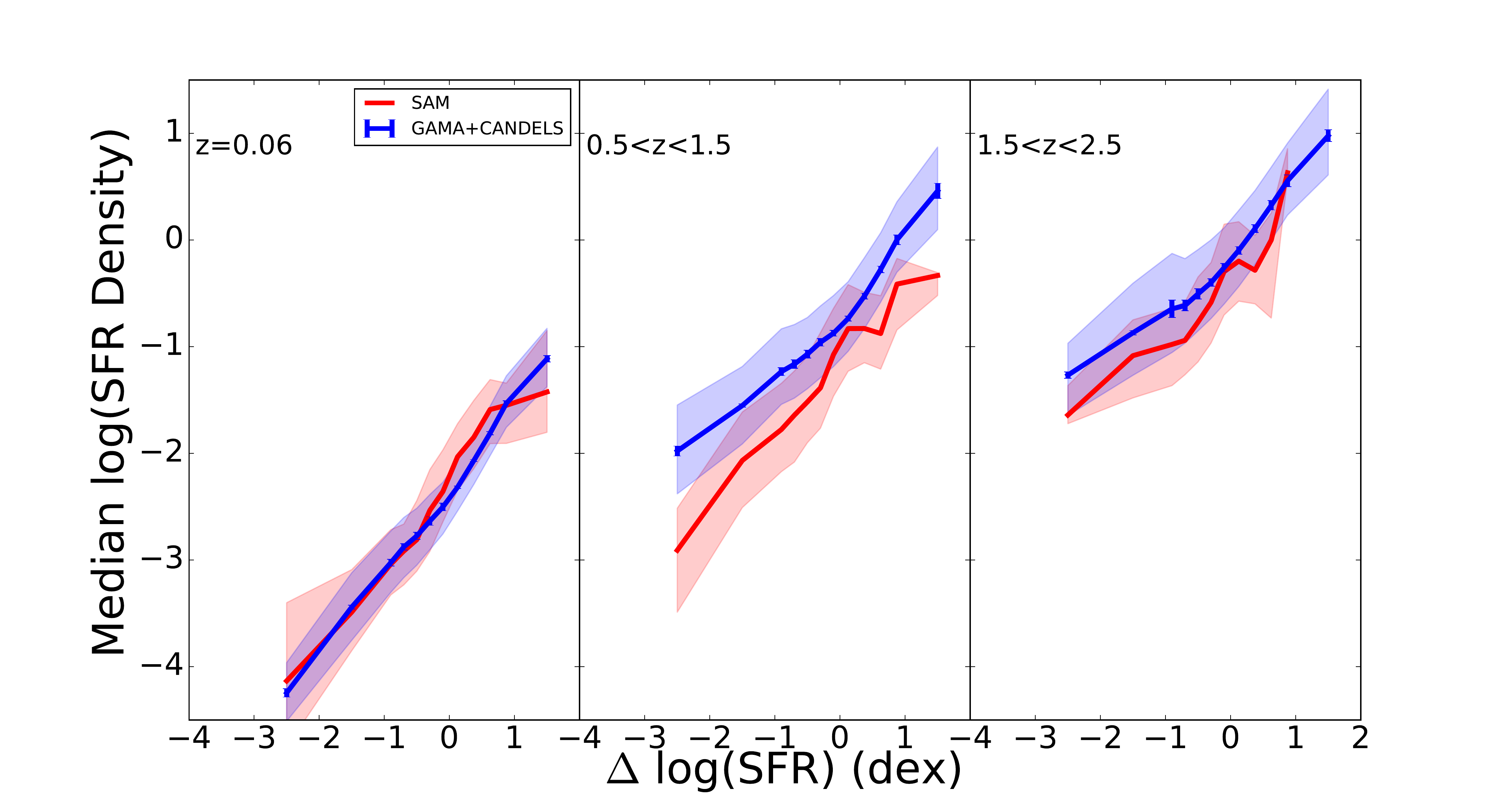, width=0.9\textwidth}
\caption{Median SFR density as a function of vertical distance from
  the fitted star-forming sequence for model galaxies (red) and
  observed galaxies (blue). The shaded region covers the 25th-75th
  percentiles of SFR density and the observations also have $1-\sigma$
  error bars reflecting the uncertainty in galaxy parameter
  estimation. The agreement between models and observations is in general quite good, although in our higher redshift bins, the models produce slightly less dense systems.
}  {\label{sigdist}}
\end{figure*}

\begin{figure*}
\epsfig{file=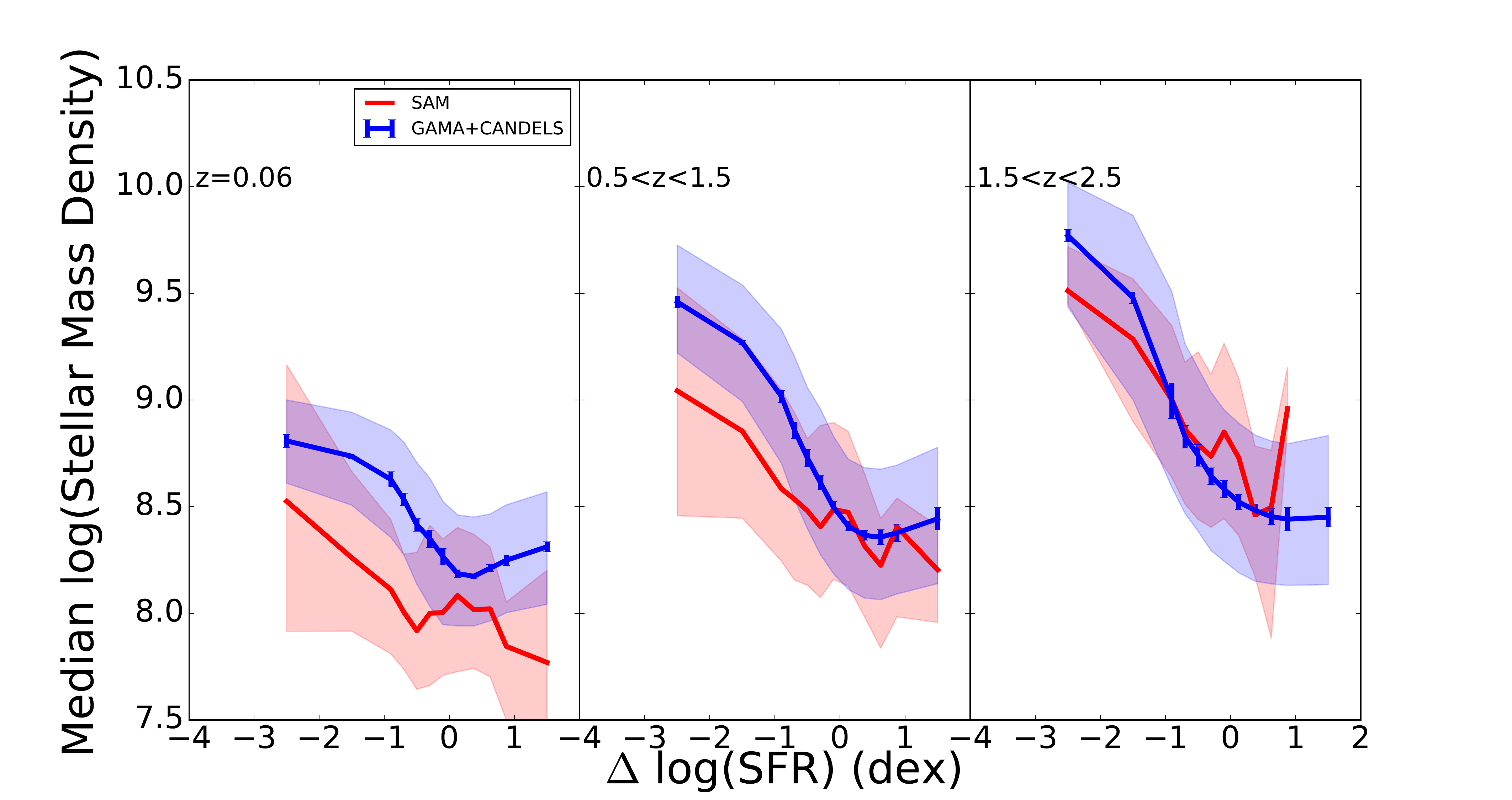, width=0.9\textwidth}
\caption{Median stellar mass density as a function of vertical
  distance from the fitted star-forming sequence for model galaxies
  (red) and observed galaxies (blue). The shaded region covers the
  25th-75th percentiles of stellar mass density and the observations
  also have $1-\sigma$ error bars reflecting uncertainty in galaxy
  parameter estimation. The agreement between the models and
  observations is generally quite good, with the largest deviation
  being $\sim0.5$ dex below the main sequence.}  {\label{sigstardist}}
\end{figure*}

Figure \ref{sigdist} shows good agreement between the median values of
$\Sigma_{\rm{SFR}}$ at all distances from the main sequence in all
three redshift bins, although model values fall slightly below observed values in our two higher redshift bins. In Figure \ref{sigstardist}, we see that the
model produces galaxies whose stellar mass surface densities are in decent agreement with those observed on the main sequence. Below the
main sequence, where the model galaxies tend to be too large, as noted
above, the stellar mass surface density falls below that found in the
observations.

\section{Distribution of Distance as a Function of Galaxy Properties}

\begin{figure*}
\epsfig{file=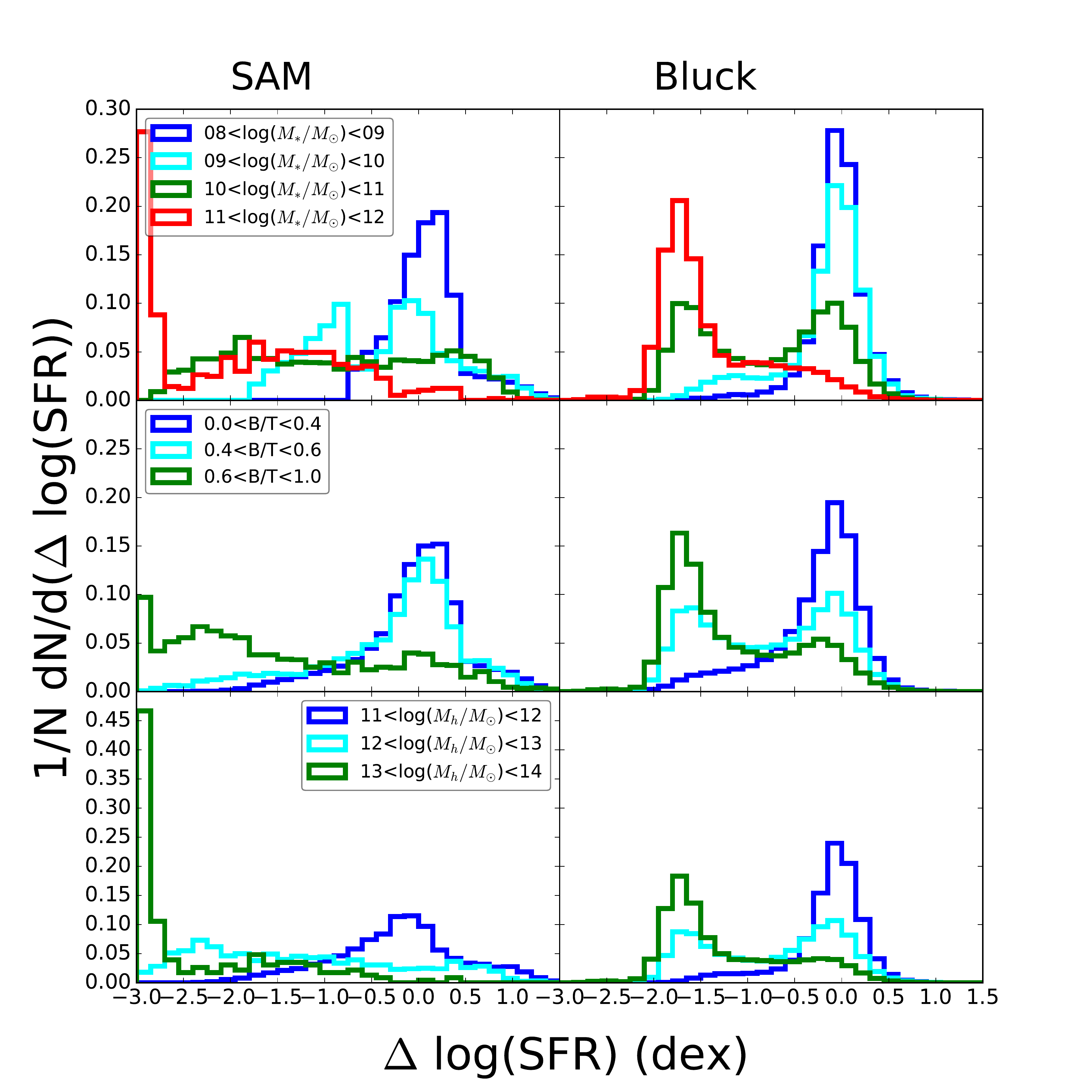, width=1.0\textwidth}
\caption{Distribution of $\Delta \log \rm{SFR}$ for different bins of galaxy properties (for galaxies with $M_{*}$>$10^{8}M_{\odot}$) in our lowest redshift bin. Left: Model properties. Right: Galaxy properties used as part of the analysis of Sloan Digital Sky Survey galaxies that span the redshift range 0.02<z<0.2 in \citet{Bluck2014}. Top panel:
  Stellar mass. Middle panel: Bulge-to-total stellar mass
  ratio (derived from bulge+disk decompositions for the observations). Bottom panel: Halo mass (derived from abundance matching for the observations). All three of these quantities behave as expected. The model and the observations qualitatively agree, although the distributions for the larger values of each galaxy parameter tend to peak farther below the main sequence in our models than in the observations.}  {\label{delsfrmass}}
\end{figure*}

Finally, we turn the tables and examine the distribution of
$\Delta$SFR in bins of various galaxy properties.  Figure
\ref{delsfrmass} shows the results for stellar mass, bulge-to-total
mass ratio, and halo mass at low redshift, for our models and for the
analysis of SDSS galaxies by \citet{Bluck2014}. The structural and stellar mass measurements for the SDSS galaxies were carried out by \citet{Simard2011} (bulge-disk decompositions by light) and \citet{Mendel2014} (bulge, disk and total stellar mass). For this plot and the next, we extend
our mass range down to $10^{8}M_{\odot}$, and use bulge-to-total
stellar mass ratio as opposed to bulge-to-total luminosity ratio, to
better compare with the results of \citet{Bluck2014}.  We see
qualitatively similar trends, with the distributions for galaxies with
larger values of these properties peaking below the main sequence. In
the top left panel, we see a very extended model distribution for high
mass galaxies. Our model distributions have less well-defined peaks
both on and off the main sequence than those seen for the observed
galaxies in the top right panel. B/T behaves the same way, although
the model peaks are a bit more well-defined (but not as well as for
the observations). The model distributions in bins of halo mass are
well stratified, with higher halo mass galaxies peaking at
successively lower $\Delta$SFR. For the highest halo mass galaxies,
this peak is right at our $\Delta$SFR floor because these are most
likely to be the most quiescent galaxies that live at the very bottom
of our SFR-$M_{*}$ plane plots. The observed high halo mass
distribution in the bottom right panel peaks at a higher $\Delta$SFR
because those galaxies live in the quiescent cloud like our quiescent
GAMA and CANDELS galaxies do. As mentioned above, the lack of a distinct peak below the main sequence in these distributions (and those throughout this section) is due to the fact that we have arbitrarily low SFRs in our model, while it becomes very difficult to measure very low SFRs observationally. In fact, for the SDSS data with which we are comparing here, an explicit floor on specific SFRs (at log(sSFR)=-12.0) has been introduced (see \citet{Brinchmann2004} for details). 

\begin{figure*}
  \epsfig{file=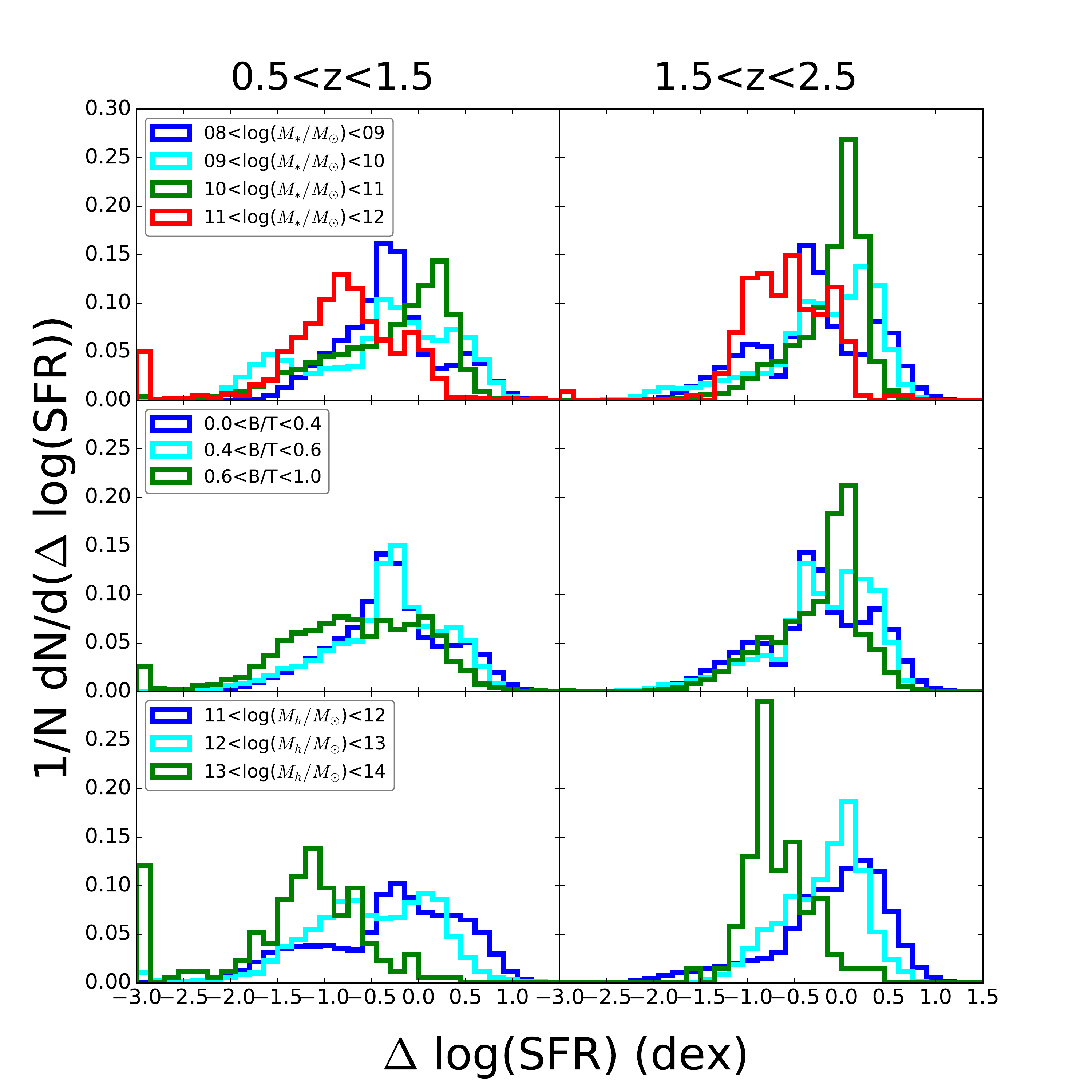, width=0.9\textwidth}
  \caption{Distribution of $\Delta \log \rm{SFR}$ for different bins of model quantities (for galaxies with $M_{*}$>$10^{8}M_{\odot}$) in our two higher redshift bins (redshift increasing left to right). Top panel: Stellar mass. Middle panel: Bulge-to-total stellar mass ratio. Bottom panel: Halo mass. We see that as we move toward higher redshift, the distributions in all bins of galaxy properties begin to pile up on the main sequence.}
  {\label{blucksam}}
\end{figure*}

In Figure \ref{blucksam}, we see the same distributions for our higher
redshift model galaxies. For all three galaxy properties, the
distributions tend to collapse onto each other as we move to higher
redshift, although stellar mass and halo mass remain somewhat
stratified.

\begin{figure*}
\epsfig{file=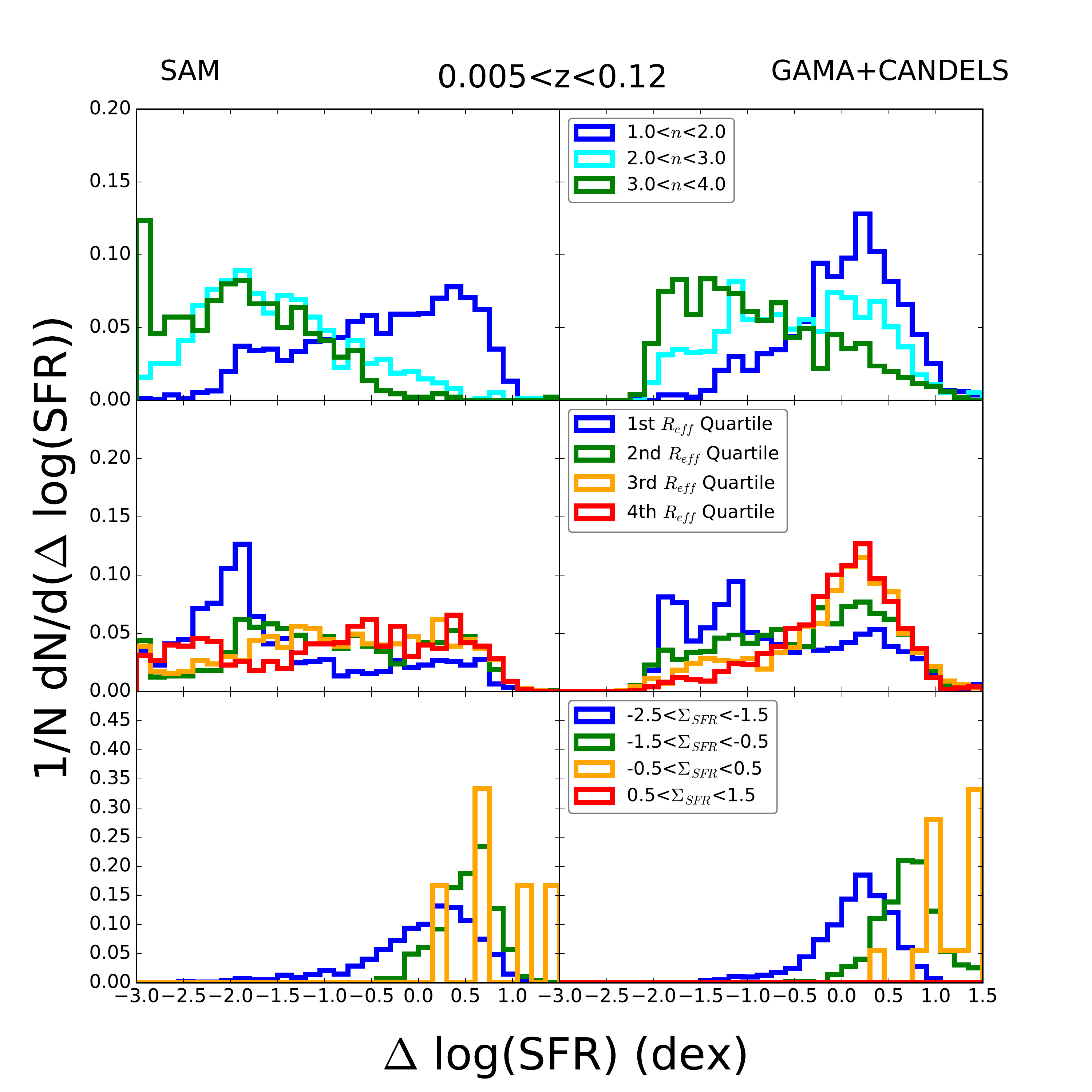, width=1.0\textwidth}
\caption{Distribution of $\Delta \log \rm{SFR}$ in our low redshift slice for
  different bins of model (left column) and observed (right column)
  galaxy properties. Top panel: S{\'e}rsic index. Middle panel:
  Quartile for effective radius for a given galaxy's 1 dex
  mass bin. Galaxies are divided into bins with
  $10^{10}$<$M_{*}$/$M_{\odot}$<$10^{11}$ and
  $10^{11}$<$M_{*}$/$M_{\odot}$<$10^{12}$. The first quartile is
  the smallest for each mass bin and so on. Bottom panel: SFR
  Density. The agreement for all three quantities is very good,
  although our model distributions tend to have tails to lower
  $\Delta$SFR than the observations.}  {\label{delsfrsersic12}}
\end{figure*}

Now we return to the quantities we have been focusing on in the
previous sections, comparing the distributions in bins of our model
quantities with those from observed galaxies. We resume using our mass
cut at $10^{10}M_{\odot}$ and revert back to using light-weighted B/T
in order to derive the model S{\'e}rsic indices in the following
plots. Figure \ref{delsfrsersic12} shows these distributions in bins
of S{\'e}rsic index, quartile of effective radius and SFR surface
density for our lowest redshift bin. To assign a radius quartile, we
divide galaxies into 1 dex mass bins ($10^{10}$-$10^{11}M_{\odot}$ and
$10^{11}$-$10^{12}M_{\odot}$) and see where they fall in the
distribution of all sizes in their respective mass bins. We see that
the qualitative agreement is good for all three quantities. Our model
distributions tend to skew to lower $\Delta$SFR as noted before. We
also see that our model doesn't stratify in radius as well as the
observations; we have some galaxies which are quite large for their
stellar mass far below the main sequence and the distribution of large
galaxies does not peak as strongly on the main sequence as it does for
observations. We also see in the bottom panels that even our most
dense star-forming systems aren't as high above the main sequence as
those seen in the observations. These conclusions are consistent with
those we reached by looking at the plots of median quantities before.

\begin{figure*}
\epsfig{file=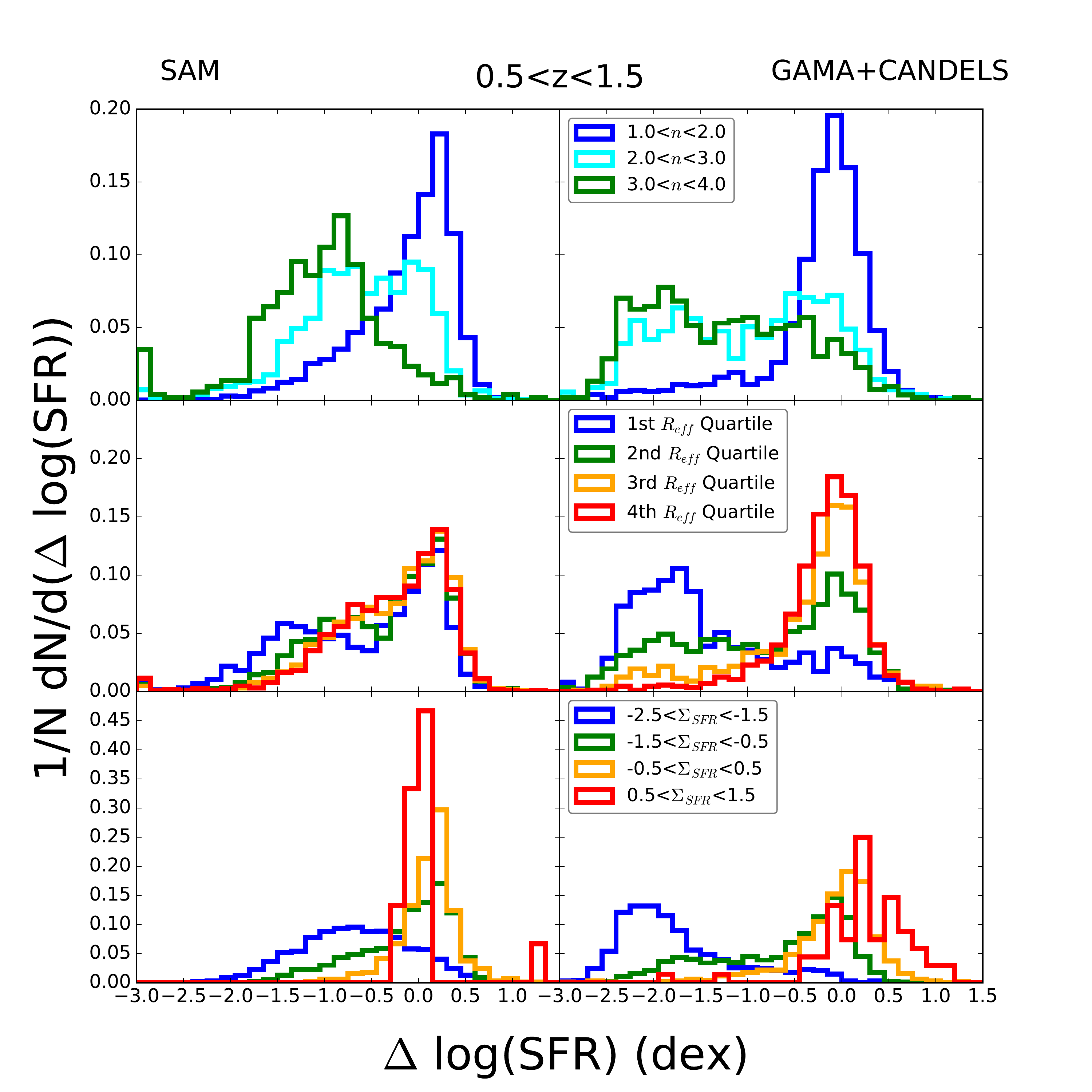, width=1.0\textwidth}
\caption{Distribution of $\Delta \log \rm{SFR}$ in our middle redshift slice for
  different bins of model (left column) and observed (right column)
  galaxy properties. Top panel: S{\'e}rsic index. Middle panel:
  Quartile for effective radius falls into for a given galaxy's 1 dex
  mass bin. Galaxies are divided into bins with
  $10^{10}$<$M_{*}$/$M_{\odot}$<$10^{11}$ and
  $10^{11}$<$M_{*}$/$M_{\odot}$<$10^{12}$. The first quartile is
  the smallest for each mass bin and so on. Bottom panel: SFR
  Density. Our agreement is again very good for S{\'e}rsic index and
  SFR density, but the models' $\Delta \log \rm{SFR}$ distributions are not as
  well differentiated for different radius quartiles the observed
  distributions. }  {\label{delsfrsersic22}}
\end{figure*}

As we move towards higher redshift, the same trends persist. Figure
\ref{delsfrsersic22} is the same as Figure \ref{delsfrsersic12} but
for our middle redshift bin. The main difference we see is in the size
distributions. While the observations show significantly different
distributions in $\Delta$SFR for the four radius quartiles, with the
most compact galaxies being much more skewed towards large negative
values of $\Delta$SFR, the $\Delta$SFR distributions in the models are
much less well separated for the different radius quartiles. A similar
result can be seen in Pandya et al. (in prep.), in which both model
and observed galaxies have been split into star-forming, transition,
and quiescent galaxies.

\begin{figure*}
\epsfig{file=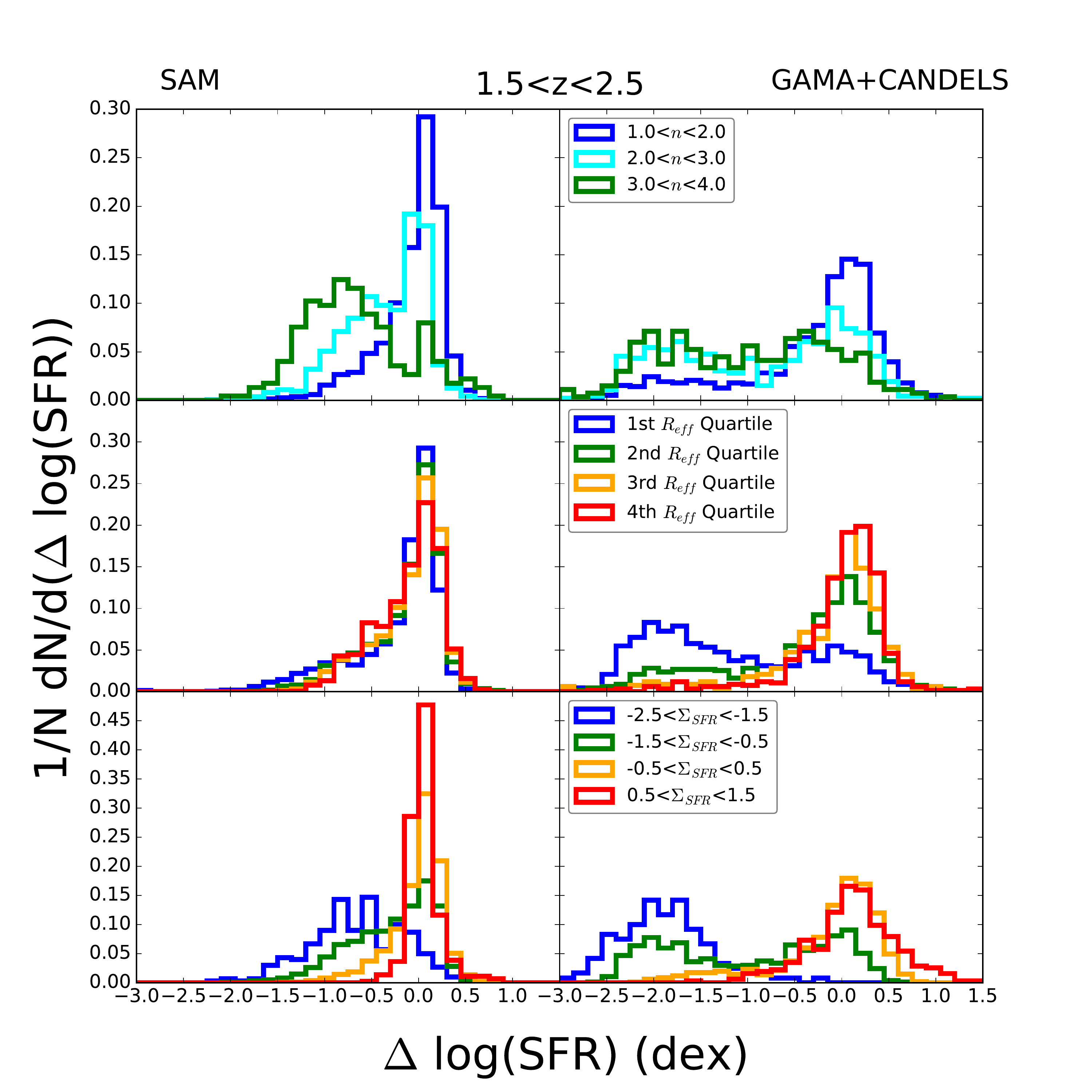, width=1.0\textwidth}
\caption{Distribution of $\Delta \log \rm{SFR}$ in our highest redshift slice
  for different bins of model (left column) and observed (right
  column) galaxy properties. Top panel: S{\'e}rsic index. Middle
  panel: Quartile for effective radius for a given galaxy's 1 dex mass
  bin. Galaxies are divided into bins with
  $10^{10}$<$M_{*}$/$M_{\odot}$<$10^{11}$ and
  $10^{11}$<$M_{*}$/$M_{\odot}$<$10^{12}$. The first quartile is
  the smallest for each mass bin and so on. Bottom panel: SFR
  Density. The main disagreement in all panels is that our model
  distributions do not have a large enough quiescent population with
  SFR well below the main sequence. }  {\label{delsfrsersic32}}
\end{figure*}

Finally, we look at high redshift in Figure
\ref{delsfrsersic32}. Here, the lack of model quiescent galaxies at
this redshift asserts itself. While our model reproduces the
separation in the distributions in bins for S{\'e}rsic index, the high
S{\'e}rsic index bins do not peak as far below the main sequence as in the
observations. This is also true for the distributions in bins of SFR
density. Our model does not reproduce the separation of distributions
in bins of size quartile, with model galaxies of all sizes living near
the main sequence. The high-S{\'e}rsic index, small-radius, and low-SFR density
peaks seen below the main sequence in the observations are the
beginnings of the quiescent cloud which our model has trouble
reproducing.

\section{Discussion}

Our study has demonstrated a significant correlation between galaxy
structural properties and their star formation activity relative to a
local star forming main sequence. These correlations have been seen
many times before both in the nearby Universe and out to high
redshift. However, our study is novel in several respects. 1) We take
particular care to carry out the analysis of the GAMA survey of nearby
galaxies and the CANDELS survey out to $z\sim 2.5$ in a consistent
manner. 2) We carry out our analysis on the WFC3 images from the full
five fields of CANDELS for the first time. 3) We make detailed
comparisons between these observations and a statistically
representative sample of model galaxies from a cosmological model of
galaxy formation and evolution. The latter point is key, as an
observed \emph{correlation} can never prove \emph{causation}, while if
we see similar correlations in models, we can at least suggest a
plausible story for a causal picture. For a short discussion
  on progenitor bias and how this might affect the causal picture, see
  Section 6.2.3.

In this section, we compare and constrast the results of our analysis
with previous results in the literature, and discuss what we have
learned about galaxy evolution in the Universe and in our models.

\subsection{Comparison with Literature Results}

\subsubsection{Comparison with the Observational Analysis of Wuyts et al.}
\label{sec:wuyts}

The study by W11 was a primary inspiration for this work, and our
observational analysis is deliberately very similar. For the most
part, our conclusions are also very similar. Here we summarize the
most important differences between the two studies. The structural
measurements used in W11 were based on the ACS I$_{814}$ image in the
1.48 deg$^2$ COSMOS field, the H$_{160}$ image in the CANDELS UDS and
GOODS-S fields, and the z$_{850}$ image in the GOODS-N
field. Similarly, the catalogs used in W11 were selected in different
filter bands and for different depths, as summarized in Table 1 of
W11. In contrast, our structural measurements are all derived from the
CANDELS H$_{160}$ image and are based on H$_{160}$-selected catalogs
with uniform depth for all five CANDELS fields. W11 computed their own
photometric redshifts and stellar masses based on compilations of
ground and space-based data from the literature for each of their
fields, while we use the CANDELS team photo-zs and stellar masses for
all five fields.

In spite of these differences, the main results of our analysis are
very much in agreement with those of W11 overall. Here we highlight a
few differences and some possible reasons for them. Comparing the
bottom row of our Figure \ref{ngrad} and Figure \ref{ndist} with Figure 1
and 2 of W11, one of the main differences we notice is the more
pronounced population of galaxies with high S{\'e}rsic index, $n\sim3.5-4$,
\emph{above} the SFMS in W11. There are two possible reasons for
this\footnote{Note also that Figure 2 and 8 of W11 plot a slightly
  different mass range than our corresponding figures.}. First, our
GALFIT quality cut eliminates galaxies with highly uncertain
S{\'e}rsic index fits. W11 did not make such a cut and so includes star
bursting systems that may have disturbed morphologies and may not be
well-fit by a single S{\'e}rsic profile.  If we remove this cut, we
also see more star bursting systems above the main sequence in our
observational sample. Secondly, the COSMOS ACS observations used by
W11 cover a much larger area than the CANDELS WFC3
footprint and includes more rare objects such as starburst galaxies
and massive, quiescent, very bulge-dominated galaxies.
Comparing our Figure \ref{regrad} and \ref{redist} with W11 Figure 3 and
8, W11 saw a slightly stronger decrease in size for galaxies above the
main sequence than we do.  Similarly, W11 see more galaxies with very
high SFR density ($\log \Sigma_{\rm SFR}>1$; Figure 4 and 8) which
again are missing from our sample. These compact, starbursting objects
are likely the same ones that we have just discussed.

\subsubsection{Comparison with other observational studies}

We find that we are in qualitative agreement with several
observational studies that have been done on the relationship between
star formation and galaxy structural properties. As shown above, we
find a similar segregation in the SFR-$M_{*}$ plane due to bulge mass
or B/T stellar mass ratio as found in \citet{Bluck2014}, the
inspiration for the plots in our Section 5. We are also in qualitative
agreement on this front with \citet{Lang2014} who found this type of
segregation in CANDELS/3D-HST data (see also a comparison with our
model in that work). \citet{Omand2014} found a simple dependence of
quiescent fraction on B/T by flipping the type of analysis done here;
they looked at the quiescent fraction in bins across the stellar
mass-bulge fraction plane and found complementary behavior to what we
have found. \citet{Wake2012}, \citet{Teimoorinia2016} and \citet{Bluck2016} all find strong dependence of star formation on central
velocity dispersion at low redshift, as we see in the left panel of
the second row of our Figure \ref{modelgrad}. \citet{Woo2013}
sees segregation in the SFR-$M_{*}$ plane due to halo mass, which we
also see quite strongly in the bottom panel of Figure
\ref{delsfrmass}. Our results are also in qualitative agreement with those of \citet{Woo2015} who examined the distribution of sSFR for galaxies in bins of $\Sigma_{\rm{1 kpc}}$ (analogous to our S{\'e}rsic index) with fixed halo mass and vice versa.

\subsubsection{Comparison with other theoretical studies}

Many studies based on semi-analytic models have shown that
bulge-dominated galaxies in massive halos tend to be red and quiescent
\citep{Bower2006, Croton2006, Somerville2008, kimm:2009, Lang2014}, in general
qualitative agreement with our results. As shown by
e.g. \citet{kimm:2009} and \citet{Lang2014}, different SAMs produce
different relative degrees of correlation of the fraction of quenched
galaxies with halo mass and bulge mass, reflecting differences in the
physical recipes responsible for quenching star formation in the
models. However, we are not aware of any other SAM-based comparison
that has examined the continuous distribution of galaxy structural
properties in the SFR-$\rm{M_{*}}$ plane as we have done here.


Such an analysis has been done for the Illustris numerical
hydrodynamical simulations at $z=0$, by \citet{Snyder2015}. The top
left panel of their Figure 5 is strikingly similar to our redshift
zero panel from Figure \ref{ngrad}, although it should be kept in mind
that their color coding is based on a different metric representing
how bulge-dominated the galaxy is (Gini-M$_{20}$). Similarly, in their
Figure 10 \citet{Snyder2015} show that quiescent galaxies are more
compact at a given mass than star forming galaxies, although they note
that the sizes of galaxies in Illustris are systematically too
large. \citet{Snyder2015} show that quenching in the Illustris
simulations is clearly associated with the growth of a massive SMBH as
well as a massive halo, very similar to what we find in our
SAMs. \citet{Snyder2015} have not presented a detailed comparison with
observations.  In another Illustris-based analysis, \citet{Sparre2015}
showed that their simulations were lacking in extreme starburst
galaxies (outliers above the SFMS), similar to what we find in our
SAMs. The predicted population of extreme starbursts is likely quite
sensitive to numerical resolution as well as the treatment of the
interstellar medium and feedback.

In another type of study, \citet{Zolotov2015} and
\citet{Tacchella2016} analyzed the star formation rates and sizes of a
set of high resolution ``zoom-in'' simulations. These 26 moderately
massive halos are not representative of a cosmological sample, and
have not been run past $z=1$, but they attain considerably higher
resolution and contain arguably more physical ``sub-grid'' recipes for
processes such as star formation and stellar feedback than large
cosmological volumes like Illustris.  \citet{Zolotov2015} and
\citet{Tacchella2016} emphasize that in their simulations, mergers and
violent disk instabilities can lead to rapid gas inflow, building a
compact, dense nucleus. They find that this ``compactification'' phase
is in general soon followed by a rapid decrease in SFR due to the
changing decreasing inflow rate relative to the stellar-driven
outflows. These authors also emphasize the role of building up a
massive halo that can support a virial shock in driving the onset of
quenching. However, these simulations do not include AGN feedback.
This is likely the reason that, as noted by these authors, galaxies in
these simulations do not ``fully quench''. It can be seen in Fig. 8 of
\citet{Tacchella2016} that the simulations contain very few galaxies
that are more than 1 dex below the SFMS, while the CANDELS
observations show a significant population of such ``strongly
quenched'' galaxies even at $1.5 < z < 2.5$. Figure 8 of
\citet{Tacchella2016} shows that within $\pm 0.5$ dex of the SFMS,
galaxies in their simulations have a weak dependence of structural
properties (stellar mass density, radius, S{\'e}rsic index) as a function of
main sequence residual. This is in qualitative agreement with our SAM
predictions, and with the CANDELS observations. It also suggests that
in order to create the strong outliers from the SFMS seen in
observations, additional physical processes (such as AGN feedback)
accompanied by fairly dramatic structural transformation are likely
needed.

\subsection{Interpretation of Results}

Overall, our model's agreement with observations is qualitatively very
good, although there are some recurring issues which manifest many
times in the above analysis. We now discuss both sides of this coin:
What does our model tell us about the Universe when the two agree with
each other, and what does the Universe tell us about our model when
they don't?

\subsubsection{What Our Model Tells Us About the Universe}
The broad agreement between our model and the observations is
extremely encouraging and suggests a plausible physical scenario that
can explain the observed correlations. In this picture, relatively
smooth accretion of gas fuels star formation and builds up
rotationally supported disks. The radial size of the disk that forms
is roughly proportional to the angular momentum of the gas, which (on
average) traces that of the dark matter halo. Relatively minor
perturbations, such as minor mergers or disk instabilities, cause
galaxies to oscillate around the SFMS as suggested in
\citet{Tacchella2016}, and seen also in our models. As long as
galaxies remain in this relatively smooth undisturbed growth phase,
their structural properties do not show a strong correlation with
their distance from the SFMS.

Eventually, either through many small perturbations or a few larger
ones (see e.g. Figure 14 of B15), a galaxy can build up a sufficiently
massive black hole that AGN feedback prevents further significant
cooling, perhaps also rapidly removing the star forming ISM through
powerful winds. In the models presented here, bulge growth and black
hole growth are explicitly linked, and both are fed through a
combination of major and minor mergers and disk instabilities. It is
certainly not clear that the details of the implementation of these
processes are correct in our simulations or any existing ones, but it
is not unexpected that the build up of a dense central nucleus and
rapid feeding of a SMBH should go together. In our models, this linked
growth of a compact, dense structure in the centers of galaxies and
the engine that drives feedback (the SMBH) is the causal driver of the
strong correlations between structure and SFR for galaxies that are
below the SFMS. It is plausible that this is also the case in the real
Universe. We note also that although there is general consensus that
what is sometimes called `halo quenching' (the build up of a halo
massive enough to sustain a virial shock) is \emph{not by itself
  sufficient to cause strong and long-duration quenching} \citep{Choi2015, Pontzen2016}, it is
certainly reasonable to suppose that any sort of AGN feedback will
have an easier time stopping the accretion of hot, low density,
isotropically distributed gas than that of dense, cold, filamentary
gas. Because the fraction of accretion via the ``hot mode'' versus
``cold mode'' increases strongly with increasing halo mass
\citep{Birnboim2003, Dekel2006, Keres2005}, it may therefore be that
some combination of halo mass and black hole mass is in fact the best
indicator of whether the conditions for quenching are met (see
Terrazas et al., in prep.; also \citet{Snyder2015} and \citet{Woo2015}). This will be an
interesting issue to explore in simulations with more detailed
treatment of AGN feedback.

Our model also suggests the existence of some rather radially large galaxies in the two lowest redshift bins most prominent in the left panel and below the main sequence in the middle panel of \ref{redist}). These galaxies do not appear to be present in existing
catalogs from GAMA or CANDELS, but an interesting question is whether
these objects could be missed due to their very low surface
brightness. The recent discovery of ``ultra diffuse galaxies'' in the
Coma and Virgo cluster
\citep{vandokkum2015a,vandokkum2015b,Koda2015,Mihos2015}, as well as
extremely large disk galaxies \citep{Ogle2015}, have called into
question whether there might be more large, diffuse galaxies out there
than previously thought.  Using the effective soft surface brightness
limit of GAMA (23.5 $\mathrm{mag/arcsec^{2}}$ in the r-band
\citep{Baldry2012}), we estimate that over 17\% of model galaxies in
our lowest redshift bin with effective radii >10 kpc at least 1.5 dex
below the main sequence would be undetected. About two-thirds of
these are disk galaxies, and the rest are spheroid dominated. A
detailed comparison between our model predictions and the observed
populations of ultra-diffuse galaxies is beyond the scope of this
paper, but it is intriguing that our models predict there may be a
population of large diffuse galaxies.

\subsubsection{What the Universe Tells Us About Our Model}

Unfortunately (or perhaps fortunately) the universe does not take our suggestions on how to run
itself, so here we discuss how our model is failing to reproduce the
observations and what we may be able to learn from this. As discussed
above, our most quenched galaxies, which are the result of intense AGN
feedback after building a massive black hole, have SFRs which are
lower than those in the quiescent cloud of observed galaxies. This
probably indicates that our treatment of AGN feedback is too
simple. Real galaxies likely undergo short duty-cycle bouts of
quenching and rejuvanation, which our simple model does not resolve.
Also, although the observed population of compact, high-central
density starburst systems well above the SFMS fits into our
theoretical merger-based picture, we have trouble actually producing
enough of these systems when compared with the observations. This is a
direct result of the treatment of star formation enhancement in
merger-triggered bursts implemented in our SAMs, which was based on a
now rather out-of-date set of hydrodynamic simulations of binary
mergers. As noted above, other hydro simulations, such as the
Illustris simulation, have had similar trouble producing ``extreme''
starbursts which we would expect to see far above the main sequence
\citep{Sparre2015}. We expect future simulations with higher
resolution and a more detailed treatment of the ISM will help us
understand how this population is produced.

The slight peaks in radial size above and below the main sequence apparent in Figure \ref{redist} appear to be due to highly star-forming galaxies
which still have a significant disk component and disk-dominated
galaxies which are slowly fading off of the main sequence,
respectively. Examples of the disky highly star-forming galaxies can
be found in the upper-rightmost occupied bins of the top middle panels
of Figures \ref{ngrad} and \ref{regrad}. These galaxies, while likely
starbursts, still have fairly low S{\'e}rsic indices and large
sizes. A few galaxies like this are enough to cause the peak seen in
Figure \ref{redist}, as we start to see fewer galaxies that far above
the main sequence overall. The large galaxies just below the main
sequence appear to be due to a slight difference in the 2D size
distributions seen in Figure \ref{regrad}. In the observational
panels, we see that in the two higher redshift bins, for a given mass,
as we move below the main sequence we only see sizes greater than or
less than the sizes seen on the main sequence. In the model, however,
at high stellar masses, especially, it is possible to encounter sizes
larger than those found on the main sequence. Because this occurs at
high mass and because the corresponding bins in Figure \ref{ngrad} are
fairly disk-dominated, it seems that these are fairly massive disks
which have fallen below the main sequence but which are not yet
quiescent. While we expect galaxies to be kept on the main sequence by
these cycles of activity and relative dormancy, it appears that
perhaps this cycle is affecting the sizes of our galaxies too
strongly, as there is no sign of this size behavior in the
observations.

As noted above, we also find that our model produces quiescent galaxies that are
somewhat larger than those observed, and this discrepancy increases
with decreasing redshift (see Figure \ref{redist}). The systematic
nature of this discrepancy in our lowest redshift bin suggests the
need to refine overall how sizes are computed.

Finally, while it is possible that some of the large model galaxies in
our lowest redshift bin might be missed observationally due to
selection effects, this is likely not the full cause of the
disagreement, especially for galaxies on the main
  sequence. Because disk galaxies are the largest galaxies for their
mass range, it would be easy to assume that most of our very large
galaxies are disk-dominated ones that escaped merging and were allowed
to grow out of control. However, more than half of our model galaxies with sizes
>20 kpc are in fact spheroid-dominated. This is doubtless due to the
limitations of the relatively simple modeling of the sizes of disks
and spheroids in our SAMs. A clue is that the sizes of our largest
galaxies, regardless of morphology, are correlated with abnormally low
halo concentrations. While the average halo concentration of our low
redshift sample is $\sim8.5$, when limiting to galaxies with effective
radii >20.0 kpc we find an average halo concentration of
$\sim7.0$. A more detailed investigation of the size-mass
  relation in our model and its evolution will be presented in
  Somerville et al. (in prep.).


\subsubsection{Progenitor Bias}

\citet{Lilly2016} have suggested that the correlations between
  star formation and structural properties might be explained by
  progenitor bias. For galaxies at any epoch, quiescent galaxies
  represent systems that left the main sequence at an earlier epoch
  when the universe was denser and galaxy sizes were
  characteristically smaller. Because of this, quiescent galaxies will
  be systematically smaller than galaxies that have continued to grow
  while on the main sequence, regardless of any relationship between
  quenching mechanism and galaxy structure. As noted above, we have
  been careful to make the distinction between correlation and
  causation in this work, but can look to our model for
  guidance. While progenitor bias exists in our model, as
  characteristic galaxy sizes grow with cosmic time, we find that we
  are unable to reproduce basic statistical galaxy properties like the
  stellar mass function, luminosity functions or stellar mass-to-halo
  mass relationship without including some form of
  feedback. Meanwhile, on the observational side, \citet{Bluck2016}
  have found that high central velocity dispersion is a good predictor
  for the fraction of green valley galaxies as well as for quiescent
  galaxies. Green valley galaxies aren't as likely to have left the
  main sequence a long time ago like quiescent galaxies, suggesting
  that feedback is a better explanation for these systems than
  progenitor bias. In light of this, while we acknowledge that
  progenitor bias may be a factor in the structural correlations
  observed here, we believe our model still represents a plausible
  explanation for our observations.

\section{Summary and Conclusions}

In this work, we have investigated the correlation of galaxy
structural properties with their location in the plane of star
formation rate and stellar mass. We studied structural properties such
as morphology as represented by S{\'e}rsic index, radial size, and
mean stellar surface density as a continuous function of a galaxy's
distance from the mean star forming main sequence at its observation
time. We carried out a parallel analysis on the GAMA survey of nearby
galaxies, the CANDELS survey which can measure galaxy structural
properties to $z\sim 3$, and a semi-analytic model that tracks the
evolution of galaxy properties within a cosmological framework. We
focus on the population of galaxies with stellar mass
>$10^{10}M_{\odot}$, for which these surveys are highly complete and
the measurement of structural properties is robust.

Our main findings are as follows:

\begin{itemize}
  \item Within $\pm 0.5$ dex of the SFMS, we find a weak dependence of
    galaxy structural properties on the distance from the MS. Below
    the main sequence, we see a rapidly steepening dependence such
    that galaxies with larger negative MS residuals had higher median
    S{\'e}rsic index, smaller size, and higher stellar surface
    density. These trends are seen in both nearby galaxies (GAMA) and
    out to $z\sim 2.5$ (CANDELS), and are qualitatively very similar
    in the theoretical models.

  \item Our observational results are very similar overall to the
    results of an earlier study by \citet[][W11]{Wuyts2011}. One
    difference between our results and those of W11 is that we do not
    find a significant population of galaxies with high S{\'e}rsic
    index ($n\sim 3.5$--4) in the extreme starburst region above the
    SFMS. Similarly, we do not see as large a population of galaxies
    with small radii above the SFMS. We suspect that these galaxies
    are removed from our sample due to our requirement of being well
    fit by a single component S{\'e}rsic profile.

  \item The good qualitative agreement between our model results and
    the observations suggests a plausible causal explanation for the
    observed correlations; namely, that central spheroids and black
    holes grow together, and black holes play a major role in
    quenching star formation in galaxies.

  \item Quantitatively, our models disagree with the observations in
    some important respects. Our models do not produce as large a
    quiescent population at high redshift ($z>1.5$) as seen in the
    observations (as already noted by B15), and the SFR for the model
    quiescent galaxies are lower than those of observed quiescent
    galaxies. This suggests the need to refine our modeling of
      AGN feedback. Moreover, the S{\'e}rsic indices of galaxies
    below the SFMS are systematically lower (more disk-like) in the
    models, while on and below the SFMS, especially at low redshift, the sizes of our galaxies are too large. As a result, there is not as large a separation between the
    sizes for the star forming and quiescent populations in the models
    as what is seen in the observations. This suggests that we
      also need to refine our determination of galaxy sizes in the
      model.
    
\end{itemize}

\section*{Acknowledgments}
We thank the anonymous referee for a constructive and
thought-provoking report. RB was supported in part by HST Theory grant
HST-AR-13270-A. rss thanks the Downsbrough family for their generous
support, and acknowledges support from the Simons Foundation through a
Simons Investigator grant. Thanks to Asa Bluck for providing us with
data, his analysis of which in his own paper was a partial inspiration
of this work. We acknowledge the contributions of hundreds of
individuals to the planning and support of the CANDELS observations,
and to the development and installation of new instruments on HST,
without which this work would not have been possible. Support for HST
Programs GO-12060 and GO-12099 was provided by NASA through grants
from the Space Telescope Science Institute, which is operated by the
Association of Universities for Research in Astronomy, Inc., under
NASA contract NAS5-26555.

\bibliographystyle{mnras}
\addcontentsline{toc}{section}{\refname}\bibliography{paper}

\end{document}